\documentclass[journal,article,submit,pdftex,moreauthors]{Definitions/mdpi} 
\usepackage{ulem} 

\usepackage[]{graphicx}
\usepackage{subcaption}

\firstpage{1} 
\pubvolume{1}
\issuenum{1}
\articlenumber{0}
\pubyear{2022}
\copyrightyear{2022}
\datereceived{} 
\dateaccepted{} 
\datepublished{} 
\hreflink{https://doi.org/} 
\pdfoutput=1

\newcommand\nat{Nature}
\newcommand\pasj{PASJ}
\newcommand\apjl{ApJL}
\newcommand\apj{ApJ}
\newcommand\aj{AJ}
\newcommand\aapr{AApR}
\newcommand\aap{AAp}

\Title{The Transformational Power of Frequency Phase Transfer Methods for ngEHT} 
\TitleCitation{ngEHT-FPT}
\AuthorCitation{Rioja et al.}

\Author{Mar\'{\i}a\,J. Rioja $^{1,2,3}$\orcidA{}, Richard Dodson $^1$\orcidB{} and Yoshiharu Asaki $^{4,5,6}$\orcidC{}}

\address{
$^1$ \quad {ICRAR, M468, The University of Western Australia, 35 Stirling Hwy, Crawley WA 6009, Australia}\\
$^2$ \quad {CSIRO Astronomy and Space Science, PO Box 1130, Bentley WA 6102, Australia}\\
$^3$ \quad{Observatorio Astron\'omico Nacional (IGN), Alfonso XII, 3 y 5, 28014 Madrid, Spain}\\
$^4$ \quad {Joint ALMA Observatory, Alonso de Córdova 3107, Vitacura, Santiago, 763 0355, Chile}\\
$^5$ \quad {National Astronomical Observatory of Japan, Alonso de C\'ordova 3788, Office 61B, Vitacura, Santiago, Chile}\\
$^6$ \quad {Department of Astronomical Science, School of Physical Sciences, The Graduate University for Advanced Studies (SOKENDAI), 2-21-1 Osawa, Mitaka, Tokyo 181-8588, Japan
}}
\corres{}
\abstract{
\typeout{\textit{Context}:}
(Sub) mm VLBI observations are strongly hindered by limited sensitivity, with the fast tropospheric fluctuations being the dominant culprit.
\typeout{\textit{Hypothesis}}
We predict great benefits from applying next-generation frequency phase transfer calibration techniques for the next generation Event Horizon Telescope, using simultaneous multi-frequency observations. 
\typeout{\textit{Method}}
We present comparative simulation studies to characterise its performance, the optimum configurations, and highlight the benefits of including observations at 85\,GHz along with the 230 and 340\,GHz bands. 
\typeout{\textit{Results}} 
The results show a transformational impact on the ngEHT array capabilities, with orders of magnitude improved sensitivity, observations routinely possible over the whole year, and ability to carry out micro-arcsecond astrometry measurements at the highest frequencies, amongst others. 
This will enable the addressing of a host of innovative open scientific questions in astrophysics.
We present a solution for highly scatter-broadened sources such as SgrA*, a prime ngEHT target.
We conclude that adding the 85\,GHz band provides a pathway to an optimum and robust performance
for ngEHT in sub-millimeter VLBI, and strongly recommmend its inclusion in the simultaneous multi-frequency receiver design.  
}


\begin{document}
\nolinenumbers  

\graphicspath{{./}{figures/}}

\section{Introduction} \label{sec:intro}
{VLBI observations at 230\,GHz  have delivered the first images of supermassive black holes.
These results highlight the unique science accessible with high observing frequencies} \citep{eht_bh,eht_sgra},
and the interest for observations at even higher frequencies and higher angular resolutions.
Nevertheless the relentless push of the upper frequency threshold in VLBI observations faces increasing challenges. 
The fast tropospheric phase fluctuations limit the length of time over which the signal can be coherently integrated (i.e. the coherence time) and prevent the application of standard phase referencing techniques applicable in the centimeter regime to extend this time.
This propagation effect poses the main challenge in VLBI observations at high frequencies, which combined with intrinsically weaker source fluxes in general, higher instrumental noise and atmospheric opacity limit the observations to the strongest target sources, 
despite many advances in imaging algorithms\citep[see][]{eht_m87_imaging}.
It also prevents astrometric measurements.

Next-generation calibration methods and technologies have the potential to overcome these limitations. 
Using trans-frequency calibration, which relies on the non-dispersive
nature of tropospheric fluctuations {(i.e. that the tropospheric phase fluctuations are proportional to observing frequency)}
has a long history \citep[e.g.][]{asaki_fpt_96, carilli_99, middelberg_05}. {It} has only fairly recently begun to deliver on its promise for mm-VLBI
\citep{rioja_11a} with the development of a suite of {strategies} hereafter grouped under the generic term `frequency phase transfer {methods}', which share a common ground in requiring multi-frequency observations. 
At heart these rely in transferring the VLBI-observable solutions (for phase, delay and rate) measured at a lower frequency to correct the residuals in the analysis of simultaneous observations of the same source at a higher frequency, after scaling the phase by the frequency ratio. 
%
Encouraged by the observational success of these techniques that have been demonstrated at frequencies up to 130\,GHz \citep[e.g.][with the KVN]{rioja_15} we propose to expand the application to ngEHT frequencies, 
where one would expect it to continue to function at two-hundred and three-hundred GHz. 
The design specifications for the ngEHT observing frequencies are under revision in the light of the possible benefits from addition of a lower frequency.
The proposed change to the system design is to add a 85\,GHz band to the original dual-frequency system at 230 and 340-GHz.
This paper is concerned with the prospects for ngEHT observations using frequency phase transfer methods and simultaneous multi-frequency receivers, with a focus on comparative performance. 
The methods are described in Section \ref{sec:meth} and the simulation studies with accurate models for atmospheric propagation effects in Section \ref{sec:aris}. 
The results and discussions in Section \ref{sec:res} present the demonstrated effectiveness, the limitations, optimal configurations and some of the enabled scientific capabilities.
These include the extension of effective coherence time at three-hundred GHz to hours, the array configuration requirements from the case study of SgrA* as a highly scatter-broadened source, 
and the astrometric measurements that could be made from such a configuration.
Section \ref{sec:conc} are the conclusions.
%
The frequency phase transfer benefits are applicable to many different science goals, so this report stands along side the partner reports on {\it Enabling transformational ngEHT science via the inclusion of 86\,GHz capabilities} {(Issaoun, S. et al. this Galaxies issue)} and {\it Applications of the source-frequency phase-referencing technique for ngEHT observations} {(Wu, J. et al. this Galaxies issue)} [references to be completed].
%

\section{The Methods}\label{sec:meth}

The so-called frequency phase transfer paradigm encompasses a family of calibration methods that have in common {the reliance of} using observations at a lower reference frequency ($\nu_{low}$) to infer the corrections at the higher target frequency ($\nu_{high}$).
Because observations at lower frequencies are more amenable, these make possible
successful outcomes where single-frequency observations at the higher frequency alone are impossible. 
Moreover, they enhance the technical and scientific capabilities of the array.

Empirical demonstrations up to 130\,GHz, detailed formulations of the methods, comprehensive error analysis, guidelines for scheduling and requirements have been presented in \citet{rioja_15,rioja_11a,rioja_20}
along with considerations for optimum performance in continuum and spectral line studies.
For example we strongly recommend an integer frequency ratio between the different frequency bands for a robust and optimum use of these techniques, but see \citet{dodson_14} for an application when the target science requires otherwise.
Here we include a brief description of these methods to support the comparative studies presented here, relevant to ngEHT. 

The Frequency Phase Transfer (hereafter FPT) calibration method relies 
on transferring the VLBI-observable solutions (for phase, delay and rate) measured at a lower frequency, $\nu_{low}$, with a temporal sampling $\tau_{low}$, 
to the analysis of simultaneous observations of the same source at a higher frequency, $\nu_{high}$,  after scaling the phase by the frequency ratio. 
This is correct to address residual non-dispersive effects (i.e. frequency independent excess path length corrections), which are precisely calibrated out at the high frequency; {these arise, for example, from unaccounted tropospheric contributions.}
On the other hand, residual dispersive effects are amplified; these arise mainly from unaccounted ionospheric and instrumental contributions (and if left uncorrected limit the functionality). 
The result of the precise tropospheric calibration is that the effective coherence time at the higher frequency is extended, allowing the detection of weaker sources than would be possible if using single-frequency observations. 
The remaining dispersive residual terms can impose limitations on the lengthened coherence time, depending on their magnitude, and additionally prevent precise astrometry. 
A FPT schedule consists of simultaneous multi-frequency observations of the target source. 
This corresponds to the observational set up used for the simulation studies presented in Section \ref{sec:aris}.

The Source/Frequency Phase Transfer (hereafter SFPR) calibration method provides a breakthrough for ultra precise mm-VLBI astrometry, beyond the scope of application of phase referencing methods, and unlimited effective coherence time.
The SFPR calibration strategy comprises two steps, the first one being the FPT described above. The second step assumes that the remaining dispersive residuals can be eliminated using interleaving observations of a second calibration source. Since those terms are slowly varying, slow telescope source switching is possible, and the sources can be widely separated. 
Using SFPR is equivalent to carrying out observations from an excellent site and with a perfect instrument; it boosts the sensitivity by reducing coherence losses, and also provides a precise astrometric registration of the images at the two frequencies. 
The SFPR comprehensive astrometric error analysis in \citet{rioja_11a} informs the astrometry estimates  presented in Section \ref{sec:mmastro}.

Multi Frequency Phase Referencing (MFPR) \citep{dodson_17} is a technique that builds on SFPR and delivers high precision astrometric measurements in the high frequency regime using  observations of the target source only. Dedicated ionospheric calibration blocks (ICE-blocks) are interleaved with the simultaneous 
pair of dual-frequency observations at mm-wavelengths;  its implementation requires an instrument with great frequency agility and wide frequency coverage, thus will be more relevant to the ngVLA. 

FPT-square \citep{zhao_17} is a technique that builds on the FPT method described above and allows for further increase of coherence time.
It uses observations of three frequency bands to form two pairs of simultaneous dual-frequency observations at mm-wavelengths and applies the scaled-correction twice to allow for the cancellation of the ionospheric contribution. As with FPT, FPT-square is not suitable for astrometric measurements in general.
%
%
%

All of these methods are widely applicable with no known upper frequency limit, if the instrument has the required {simultaneous multi-frequency} capability. 
These have been demonstrated for continuum and spectral line VLBI with ground observations up to 130\,GHz, and at integer and non-integer frequency ratios. 
Furthermore they are potentially applicable {in the} space VLBI {domain} \citep{rioja_11b},
{improving the outcomes by calibrating out satellite orbit errors which have a non-dispersive nature, just as for the tropospheric errors}.

\section{Simulations for Coherence and Astrometric Studies at ngEHT frequencies}\label{sec:aris}

We have carried out comparative simulation studies along with astrometric error propagation to quantify the benefits of FPT and SFPR multi-frequency techniques  applied to ngEHT observing frequencies (ie bands around 85, 230 and 340\,GHz).
We focus on their power to overcome the dominant challenge imposed by the fast tropospheric phase fluctuations, which severely limit the coherence time and prevent astrometric measurements. 



We have used ARIS \citep{a07} as our simulation tool\footnote{github.com/yoshi-asaki/aris} to generate synthetic datasets. 
ARIS was designed explicitly for astrometric studies and includes realistic semi-analytical models of the so-called static and dynamic terms to mimic the troposphere and the ionosphere atmospheric propagation effects. 
The latter are implemented as moving phase screens assuming Kolmogorov turbulence characterized by a constant scale factor given by the C$_w$ coefficient, with higher values indicative of stronger fluctuations, as expected from worsening weather conditions and/or lower quality sites.
The output comprises two files, one for the target and another for the reference dataset.
These can refer to observations along different lines of sight at the same frequency, as in phase referencing, or to simultaneous observations at different frequencies along the same line of sight, for the multi-frequency studies presented here. 
The output data files are in IDI-FITS format suitable for processing in most data reduction packages.

The studies presented here focus on the dominant atmospheric propagation effects. The simulations include four antenna sites, namely Owens Valley, Mauna Kea, KVN Yonsei and the Large Millimeter Telescope, which 
result in a range of baselines up to 6,000\,km.
The target was selected to be a point source at a declination of +59$^o$, allowing tracking for hours with Zenith angles less than 70$^o$ for all antennas. 
The simulations are for the strong signal regime, that is they do not include thermal noise.
The atmospheric weather is imposed without regard for the nominal site, that is the actual antenna locations play an arbitrary role.
%
The explored parameter space of atmospheric propagation effects are simulated 
using a dynamic turbulent tropospheric phase screen that is scaled by a range of $C_w$ values equal to 0.5, 1 and 2. Best weather conditions are dubbed V for Very Good, with C$_w$ of 0.5 and doubling for so called Good (dubbed G), and Tolerable (dubbed T) weather conditions. 
The static contributions use randomly determined values with standard deviations for the tropospheric excess path delay (i.e. $\Delta \ell$) of 3~cm and the ionospheric residual electron content (i.e. $\Delta$TEC) of 6~TECU, at each VLBI station.
The set of observing frequencies correspond to the originally planned 230 and 340\,GHz bands, plus the new proposed 85 and 255\,GHz bands. The latter is for testing the impact of frequency ratios being integer or not.
We generated a complete set of synthetic single-frequency and simultaneous dual-frequency datasets, for all frequency and frequency pairs, and for all weather conditions.

The analysis of all the datasets was carried out with AIPS, using FPT calibration techniques for the dual-frequency datasets. It comprised of a self-calibration  at the reference frequency ($\nu_{low}$) with a range of solution intervals given by $\tau_{low}$ equal to 8, 15, 30 and 60 seconds, which are used to stabilise the phases at the target frequency $\nu_{high}$. {This} was followed up with a second calibration at the target frequency with a much longer solution interval, given by $\tau_{high}$ equal to 1, 3, 10 and 60\,minutes.
These are very different scales as $\tau_{low}$ {is related} to the atmospheric-coherence time and $\tau_{high}$ corresponds to the FPT-coherence time, {that is, after conditioning with FPT calibration.}
For the single-frequency datasets we use standard calibration.
In all cases, following calibration, we measured the peak flux density in the images.
Our preferred approach to calculate the coherence is to use 
the fractional peak flux recovery (FFR) quantity, calculated from the peak flux density value divided by the model point source intensity, as indicative of the coherence losses resulting from the calibration strategy. 
Another approach consists of measuring the baseline coherence over various timescales (e.g. using the AIPS task UVRMS).The former is closer, we feel, to what is required to characterize the quality of the image recovery. For example the baseline coherence can be 100\% whilst the FFR is zero, if the baselines are out of phase. Thus all figures on coherence measurements use FFR, {except} for Figure \ref{fig:coh_uvrms}.



Finally, we use the SFPR astrometric propagation error formulation in \citet{rioja_11a} to calculate the  accuracy in astrometric measurements enabled with the simultaneous dual-frequency observations at ngEHT frequencies.

\section{Results and Discussions}\label{sec:res}

\subsection{Verification of weather models in ARIS}\label{sec:raris}

\begin{figure}
    \centering
    \includegraphics[width=0.48\textwidth]{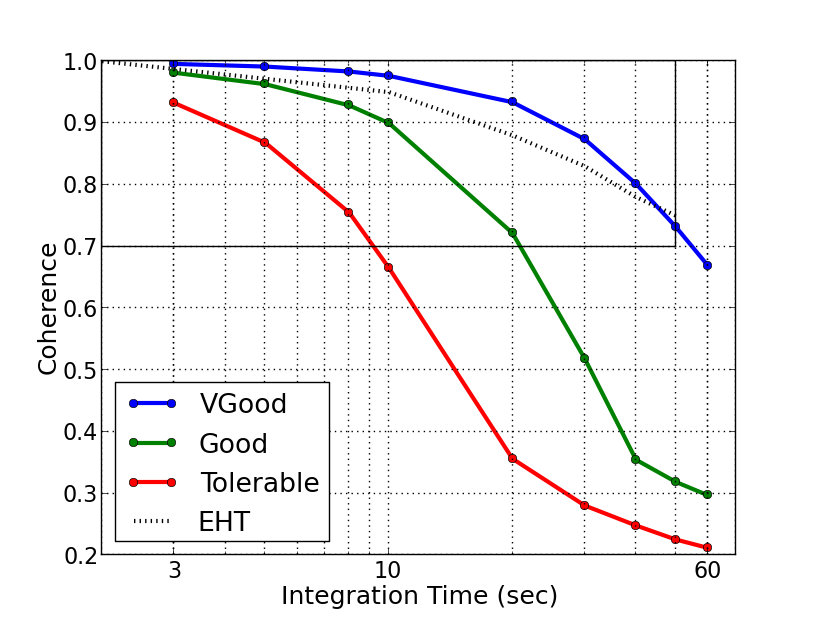}
    \includegraphics[width=0.45\textwidth]{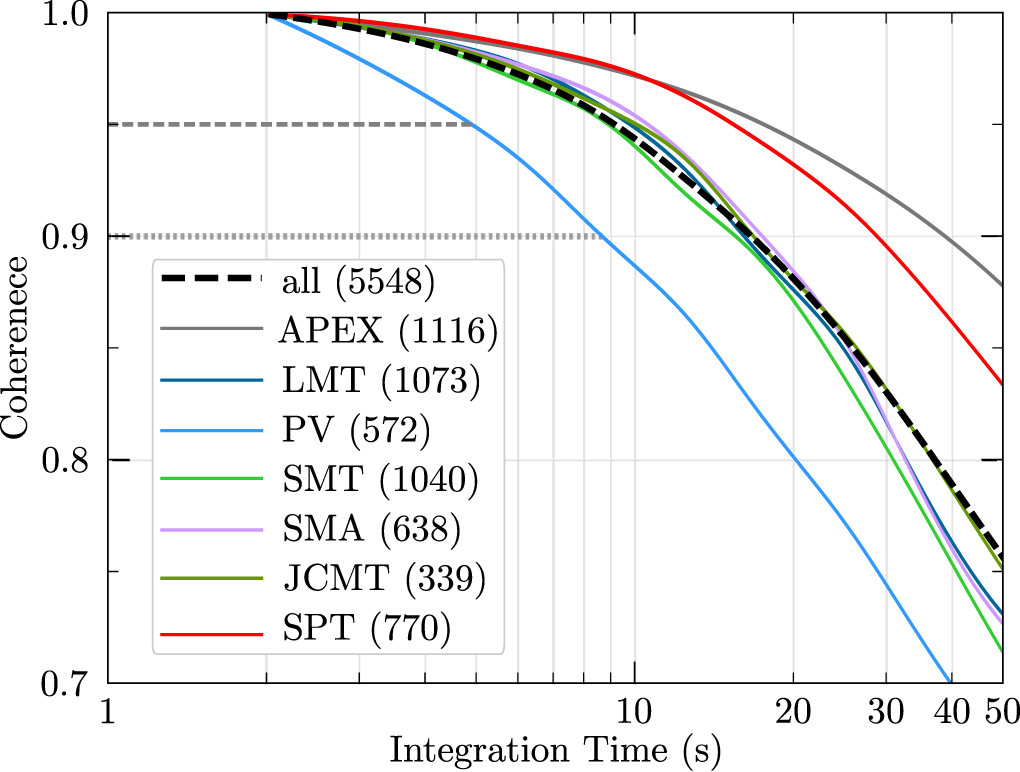}
    \caption{
    Comparison between atmospheric coherence measured from ARIS simulations and the empirical EHT datasets for a range of integration times up to 50 seconds, at 230\,GHz. Plotted are first quartile values averaged over all baselines.
    {\it Left}: Solid lines are for the synthetic datasets under three weather conditions: Very Good (blue, C$_w$=0.5), Good (green, C$_w$=1) and Tolerable (green, C$_w$=2).  The black dotted line is for empirical EHT observations. The simulations for Very Good weather conditions (i.e typical ALMA conditions) reproduce well the empirical measurements.
    {\it Right}: Fig. 2 from \citet{m87_eht_II} shows EHT empirical measurements. 
    Solid lines are  for individual baselines to ALMA and the dotted line is the baseline averaged coherence; the latter is  overplotted in left. The axis limits are shown on the left as light black lines. 
    \label{fig:coh_uvrms}}
\end{figure}

{We use the comparison between outcomes from the empirical EHT observations at 230\,GHz and our synthetic datasets, to confirm the correspondence of the ARIS tropospheric models with the real weather conditions at EHT sites.  
Figure \ref{fig:coh_uvrms} (left) shows the 1st quartile baseline atmospheric coherence calculated with the AIPS task UVRMS as a function of integration time for our simulations under a range of weather conditions, shown with solid lines with different colors. 
Overplotted with a dotted line are the results measured from EHT observations as shown in the EHT data paper \citep[]{m87_eht_II}.  
We conclude that the EHT empirical results match those of Very Good weather in our simulations, which corresponds to typical weather conditions at the ALMA site (without WVR corrections).  
This step is a fundamental check of the validity of the simulation studies and the conclusions extracted.
}

\subsection{FPT Coherence at 340\,GHz using 230\,GHz as the reference frequency}\label{sec:230} 

Here we present the prospects for the original system design which encompasses a dual-frequency receiver covering the 230 and 340-GHz bands.

\begin{figure}
    \centering
    \includegraphics[width=0.48\textwidth]{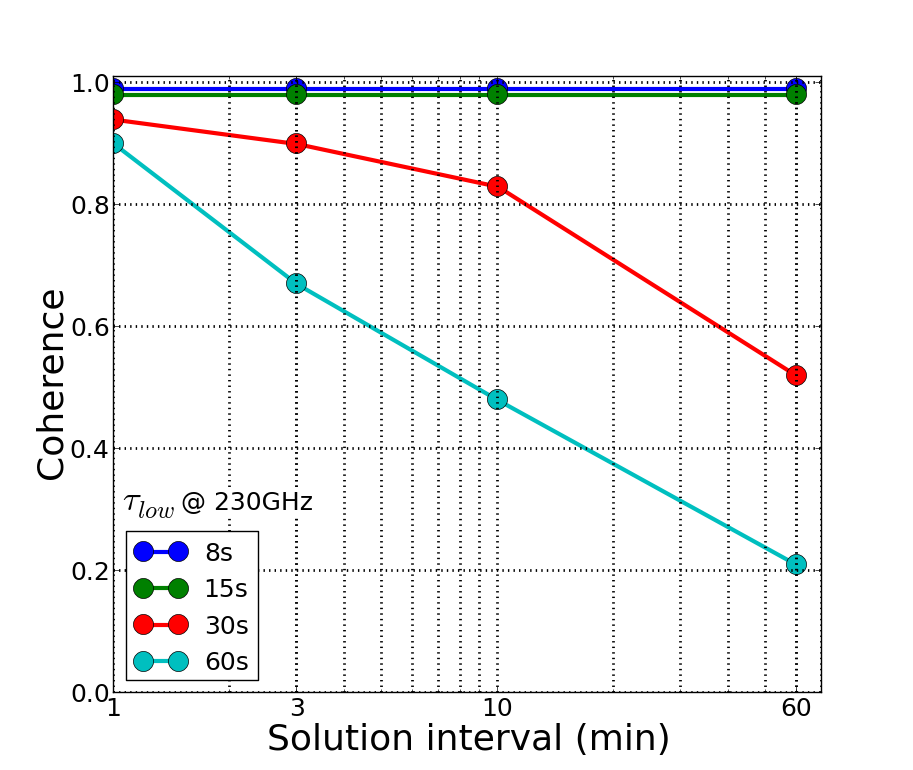}
    \includegraphics[width=0.48\textwidth]{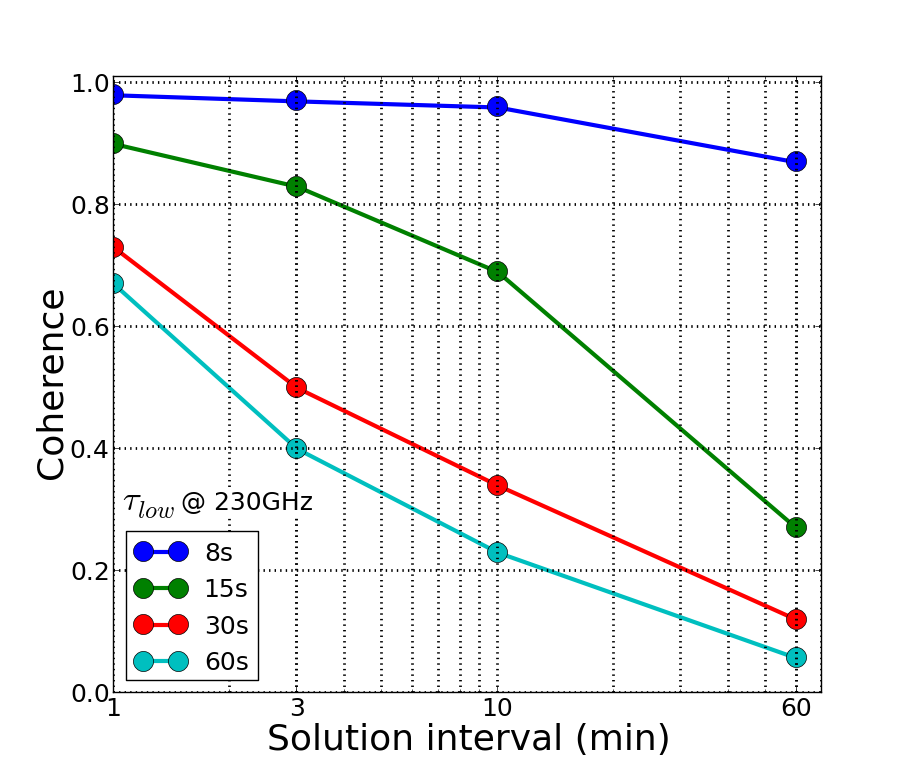}
    \caption{
    Expectations for FPT Coherence for the 340\,GHz  band under Very Good ({\it Left}) and Good ({\it Right})  weather conditions, after calibration with simultaneous 230\,GHz band observations, for a range of integration times up to 60 minutes. 
    Different colors are for different calibration timescales at 230\,GHz (i.e. $\tau_{low}$) as listed in the inset: 8 seconds (dark blue), 15 seconds (green), 30 seconds (red) and 60 seconds (cyan). 
    In Very Good weather $\tau_{low} \le$15 sec results in sustained high coherence ($\gg$0.9); longer $\tau_{low}$ timescales leads to degradation of the coherence, dropping to a coherence of 0.9 at 340\,GHz  for solution intervals up to 3 minutes or 0.8 for 10 minute solutions intervals. This degradation is caused by the non integer frequency ratio in these simulations. 
Under G weather conditions with $\tau_{low}\sim$8\,sec we achieve coherence of 0.9 at 340\,GHz  up to an hour.
    \label{fig:coh_230}}
\end{figure}

Figure \ref{fig:coh_230} plots the FPT coherence at 340\,GHz ($\nu_{high}$) as a function of calibration timescales ($\tau_{high}$), for a range of values between 1 and 60 minutes, {after FPT calibration using simultaneous 230\,GHz ($\nu_{low}$) observations. Shown are performances} 
under Very Good (V, {\it left}) and Good (G, {\it right}) weather conditions. 
Different colors correspond to different calibration timescales at 230\,GHz ($\tau_{low}$), ranging between 8 and 60\,seconds. The coherence corresponds to the FFR quantity obtained as described in Section \ref{sec:aris}.
We find similar results for the 255-GHz and 340-GHz frequency pair.

Under V weather conditions the results show very high coherence $\gg$0.9 at 340\,GHz, for solution intervals up to an hour and beyond, using $\tau_{low} \le$ 15 seconds.
Longer $\tau_{low}$, $\sim$30\,seconds, achieves a coherence of 0.9 at 340\,GHz  for solution intervals  $\tau_{high}$ up to 3 minutes; longer solution intervals result in larger coherence losses (e.g. 10 minutes results in 20\% loss). 
%
The performance deteriorates under G weather, but high coherence of $\gg$0.9 at 340\,GHz can be obtained for up to 10\,minutes with a sufficiently short $\tau_{low}$ of $\sim$8\,seconds.
The degradation of performance is mainly due to the non-integer frequency ratio ($\nu_{high}\over \nu_{low}$), as discussed in Section \ref{sec:nonint}.

We conclude that using 230-GHz as the reference frequency and under Very Good  weather conditions leads to acceptable results at 340-GHz. Furthermore acceptable results are possible under Good weather conditions, with the more stringent constraint that $\tau_{low}$ is sufficiently fast.
We remind the reader that for FPT a direct detection at $\nu_{low}$ within $\tau_{low}$ is a fundamental requirement.
With $\nu_{low}$ equal to 230\,GHz this will limit the number of possible targets. 
For example, only approximately 20\% of the ALMA calibrator source list would be detectable by ngEHT with a $\tau_{low}$ of 8\,seconds. 
This is, of course, significantly better than the percentage of that list which would be directly detected with single frequency observations at 340-GHz, which is about 8\%.

\subsection{FPT Coherence at 340\,GHz using 85\,GHz as the reference frequency}\label{sec:85}

\begin{figure}
    \centering
    \includegraphics[width=0.48\textwidth]{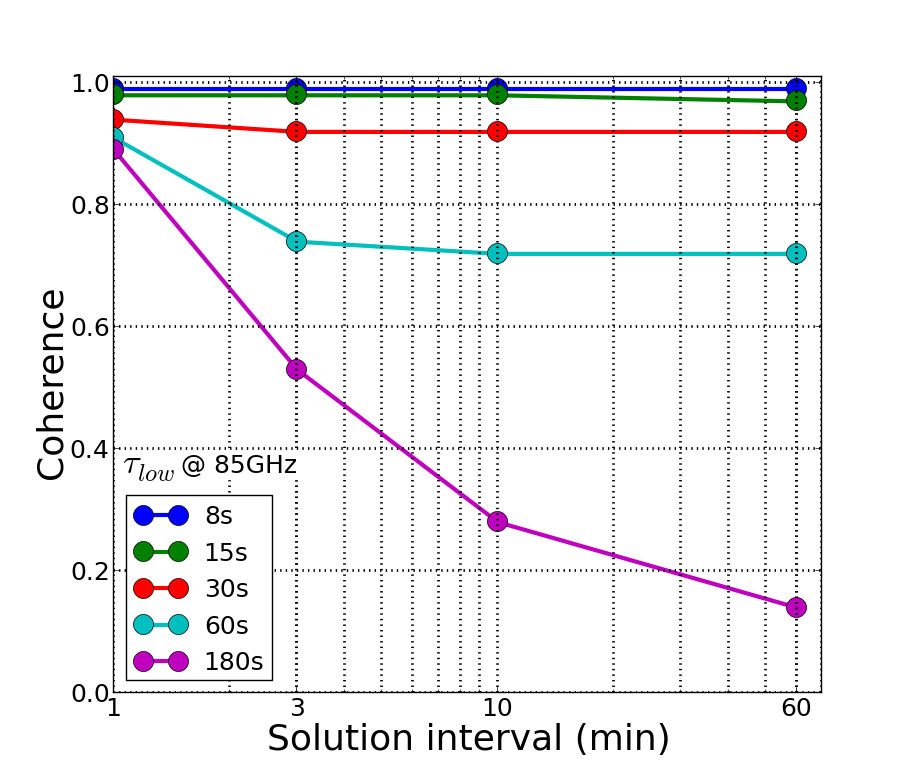}
    \includegraphics[width=0.48\textwidth]{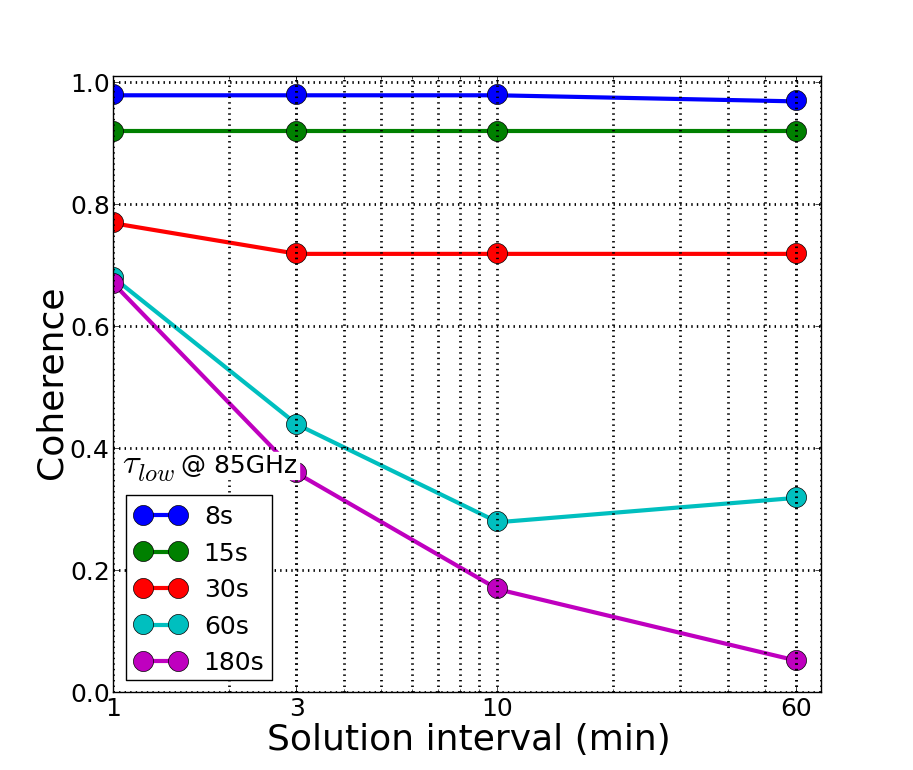}
    \includegraphics[width=0.48\textwidth]{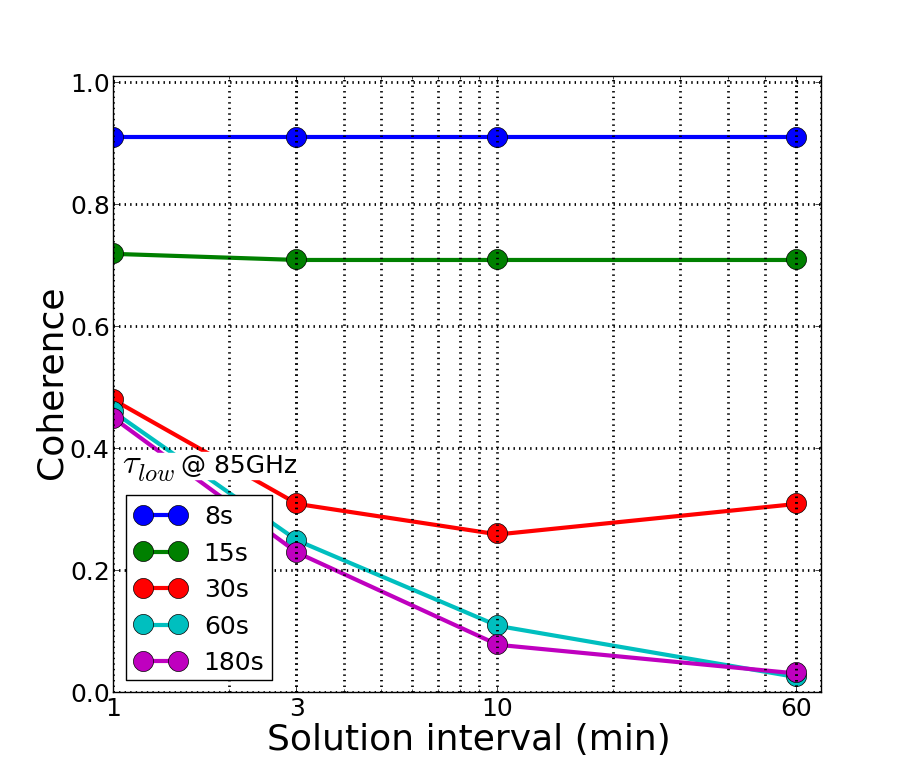}
    \caption{\label{fig:coh_85} 
    Expectations for FPT Coherence at 340\,GHz  band under Very Good ({\it left}) and Good ({\it right}) and Tolerable ({\it bottom}) weather conditions, after calibration with simultaneous {85}\,GHz band observations, for a range of integration times up to 60 minutes. 
    Different colors are for different calibration timescales at {85}\,GHz (i.e. $\tau_{low}$) as listed in the inset: 8 seconds (dark blue), 15 seconds (green), 30 seconds (red), 60 seconds (cyan) and 180 seconds (magenta). 
    Sustained coherence of 0.9 or better can be obtained with calibration timescales of 30, 15 and 8 seconds, respectively. 
    In very good weather coherence of 0.7 is obtained even with a 1\,min calibration cycle, and similar levels of coherence can be achieved with a 30\,s calibration cycle in good weather. Even in tolerable weather similar coherence is achievable with a fast 15\,second $\tau_{low}$.
    }
\end{figure}

%
With the reference frequency at 85\,GHz, the target frequencies at 255- and 340-GHz can be integer frequency ratios (3 and 4 respectively), thus 
the solutions are more robust and coherence is not lost.
Figure \ref{fig:coh_85} plots the FPT coherence at 340\,GHz ($\nu_{high}$) as a function of calibration timescales ($\tau_{high}$), for a range of values between 1 and 60 minutes, under Very Good (V, {\it left}), Good (G, {\it right}) and Tolerable (T, {\it bottom}) weather conditions, after FPT calibration using simultaneous 85\,GHz ($\nu_{low}$) observations. Different colors correspond to different calibration timescales ($\tau_{low}$) at 85\,GHz, ranging between 8 and 180\,seconds.

Under V, G and T weather conditions we obtain very high coherence $\gg$0.9 at 340\,GHz, for solution intervals up to an hour and beyond, using $\tau_{low} \le$ 30, 15 and 8\,seconds respectively.
Even with $\tau_{low}$ double these limits, 
long term coherence greater than 0.7 is maintained for an hour and beyond, as the frequency ratio is integer and phase ambiguity issues have no impact. 

We conclude that using 85-GHz as the reference frequency and in Very Good and Good  weather conditions, superior results are possible at 340-GHz, and with longer calibration time scales compared to using 230-GHz as the reference. 
%
Furthermore, acceptable results are possible under Tolerable weather conditions, with the more stringent constraint that $\tau_{low}$ is sufficiently fast.
Approximately 90\% of the ALMA calibrator source list would be detectable by ngEHT at 85\,GHz with $\tau_{low}$ of 15\,seconds.
Thus using 85-GHz as the reference frequency will dramatically increase the number of viable targets, allowing for demographic studies at 340-GHz.

\begin{figure}
    \centering
    \includegraphics[width=0.6\textwidth]{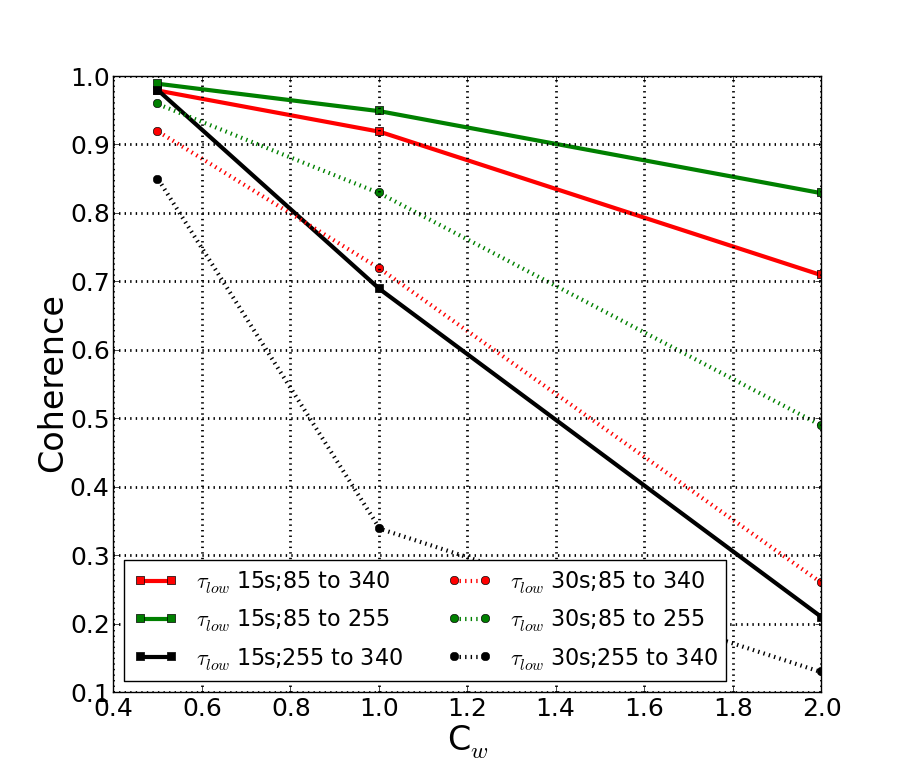}
    \caption{
    Expectations for FPT Coherence at 10 minutes integration time for a range of weather conditions (Very Good to Tolerable).
    Different colors are for all combinations of frequency pairs: green and red for pairs with {85}\,GHz as the reference frequency (i.e. 85$\rightarrow$255 and 85$\rightarrow$340 respectively) and black for 255\,GHz as the reference frequency (i.e. 255$\rightarrow$340). 
    Solid and dashed lines are for calibration timescales at the low frequency equal to 15 seconds and 30 seconds, respectively.
    Higher frequencies and worse weather increases the degradation, but faster $\tau_{low}$ and integer frequency ratios improves the performance.
    \label{fig:coh_cw}}
\end{figure}

Figure \ref{fig:coh_cw} shows a compendia of the results on FPT coherence for $\tau_{high}$ of 10\,minutes for three frequency pairs (85$\rightarrow$340\,GHz, 85$\rightarrow$255\,GHz and 255$\rightarrow$340\,GHz) and two $\tau_{low}$ values (15\,s and 30\,s), as a function of the weather conditions, given by the $C_w$ parameter (i.e. values 0.5, 1 and 2 for V, G and T respectively).
%
%
Figure \ref{fig:coh_cw} allows us to draw some general conclusions: {\it a}) worsening weather conditions lead to degradation in the phase coherence, {\it b}) that, for a given pair of frequencies, the coherence gets worse faster as $\tau_{low}$ is longer, {\it c}) that, for a given $\nu_{low}$, the higher $\nu_{high}$ the faster the degradation and {\it d}) that fastest degradation is for a non-integer frequency ratio.

%
FPT can achieve increased effective sensitivity at (sub)mm-VLBI because it enables extended integration times, which allows the detection of sources that were otherwise too weak to be detected within the atmospheric coherence time with single-frequency observations.
Our results suggest that the best performance at 340-GHz is achieved using 85-GHz as the reference frequency, compared to 230-GHz.
We conclude this based on the following reasons. 
This approach provides superior calibration at 340-GHz, that is higher coherence and for longer integration times. 
Additionally, observations at 340- referenced to 85\,GHz will be more robust, as they will work under a wider range of weather conditions 
massively expanding the observational window 
and/or increasing the number of suitable sites. This is partially because the analysis of data with integer frequency ratios is more straight-forward.
Furthermore, direct detections at 85\,GHz will be both more numerous and have higher SNR, due to the sources being intrinsically brighter, system noise being lower and $\tau_{low}$ being longer. This translates into  wide  applicability as more targets will be observable. 
Finally, 85-GHz provides a path to ultra-precise relative astrometry at 340-GHz, as discussed in Section \ref{sec:mmastro}. 

{Using a lower reference frequency, for example 43 GHz, results in doubling the scaling factor (given by the frequency ratio) and therefore increasing the propagation of the phase observable errors to the higher frequency. Additionally, the residual FPT ionospheric errors are larger using 43\,GHz, compared to 85\,GHz (see \citep{rioja_11a} for details on the impact of the frequency ratio). Lastly, 85\,GHz band offers important benefits for observations of SgrA*, a main EHT target, as discussed in section \ref{sec:sgra}.  We conclude that 85-GHz is the reference frequency of choice for ngEHT.}

\begin{figure}
    \centering
    \includegraphics[width=0.6\textwidth]{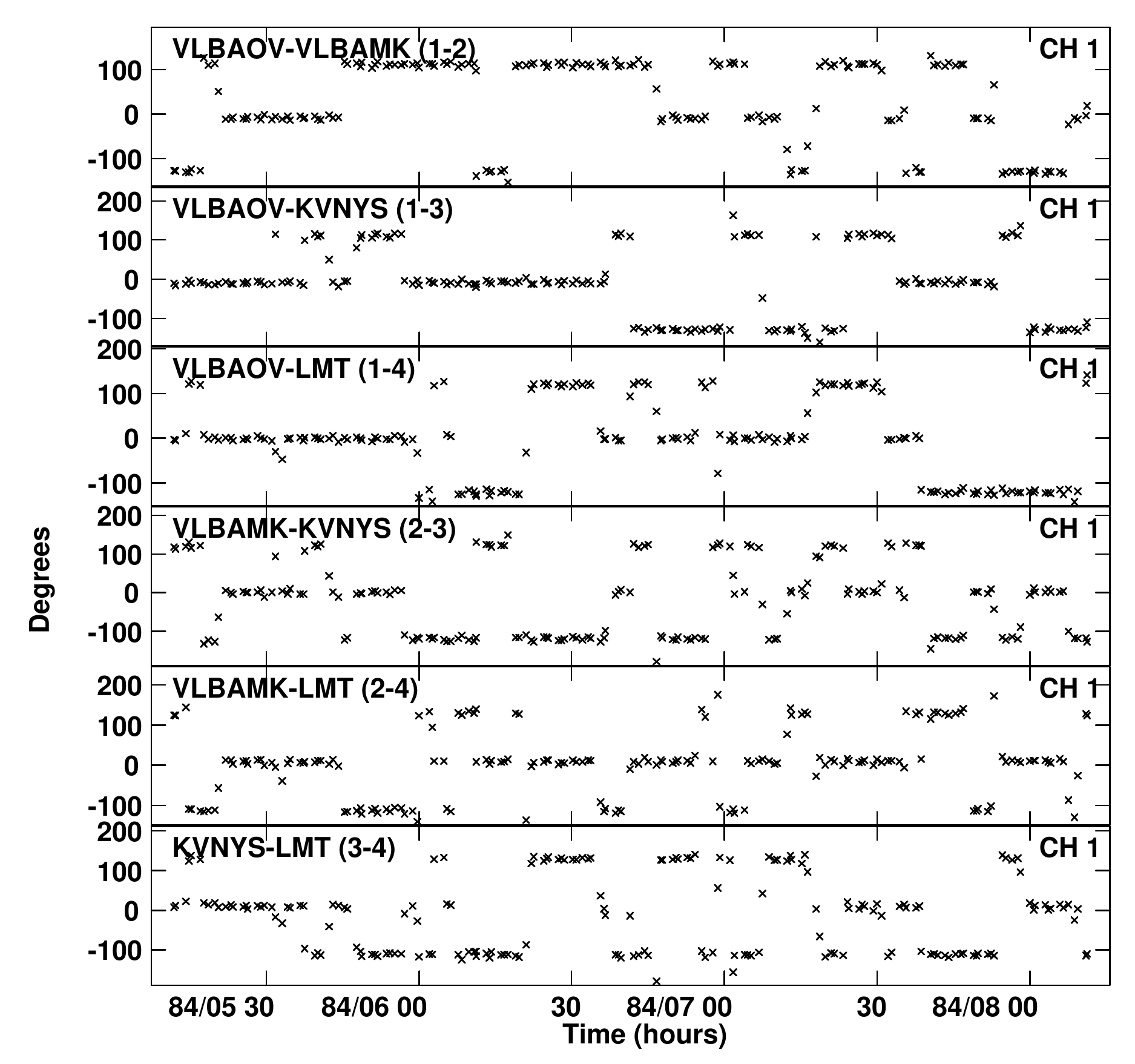}
    \caption{
    Impact of a non-integer frequency ratio between different observing bands. Here shown are FPT residual visibility phases at 340\,GHz after calibration with simultaneous 255\,GHz band observations and a calibration timescale equal to 15 seconds, under Good weather conditions. Phase jumps are a consequence of occasionally failing to track ambiguities. When an ambiguity is lost a phase jump of $\pm$120$^o$ is introduced. 
    \label{fig:nonint}}
\end{figure}

\subsection{Impact of non-integer frequency ratios}\label{sec:nonint}

In frequency phase transfer-based methods the frequency ratio is used to scale the phase measured at the low frequency. 
Therefore, for a non-integer frequency ratio the failure to track {intrinsic $2\pi$} phase ambiguities introduces offsets and jumps in the calibrated phases at the high frequency. These jumps have a significant impact in the FPT-coherence, 
{namely, that the phase does not change smoothly, but falls on multiple levels. }
Figure \ref{fig:nonint} shows the synthetic thermal-noise free {residual visibility phases at 340\,GHz} after {FPT} calibration with 255\,GHz  simultaneous observations.
Here the frequency ratio is 340/255 and thus the phase jump is 
{MOD(340/255,1) i.e. exactly 1/3 of a turn, or }
$\pm$120$^o$. Self-calibration on this data on longer timescales will introduce losses, as one would average over multiple levels. 
These losses are avoided if the ratio is integer, so that no phase jump is introduced, or in the non-integer case if no ambiguities are lost, which can be achieved under Very Good weather conditions (i.e. C$_w\ll$1) or with very short solution intervals on the lower frequency $(\tau_{low}$).

We note that observations are performed over wide bandwidths, rather than at a single frequency. The frequency reference points are selectable and these are all that need to have an integer ratio; these can in principle be outside the observed bandwidth \citep[e.g.][]{dodson_14}. We would not recommend extrapolating too far from the observed frequency band to reach the reference point; we suggest no further than the spanned bandwidth.
We strongly recommend using integer frequency ratios that enable a robust performance and applicability, optimizing the coherence achieved under all weather conditions, allowing for increased sensitivity because of the possibility of longer $\tau_{high}$ and enabling unambiguous bona-fide astrometry. 

\subsection{Simulations of  SgrA* observations, a Case Study for Scattered Sources} \label{sec:sgra}

\begin{figure}
    \centering
    \centering
    \includegraphics[width=0.6\textwidth]{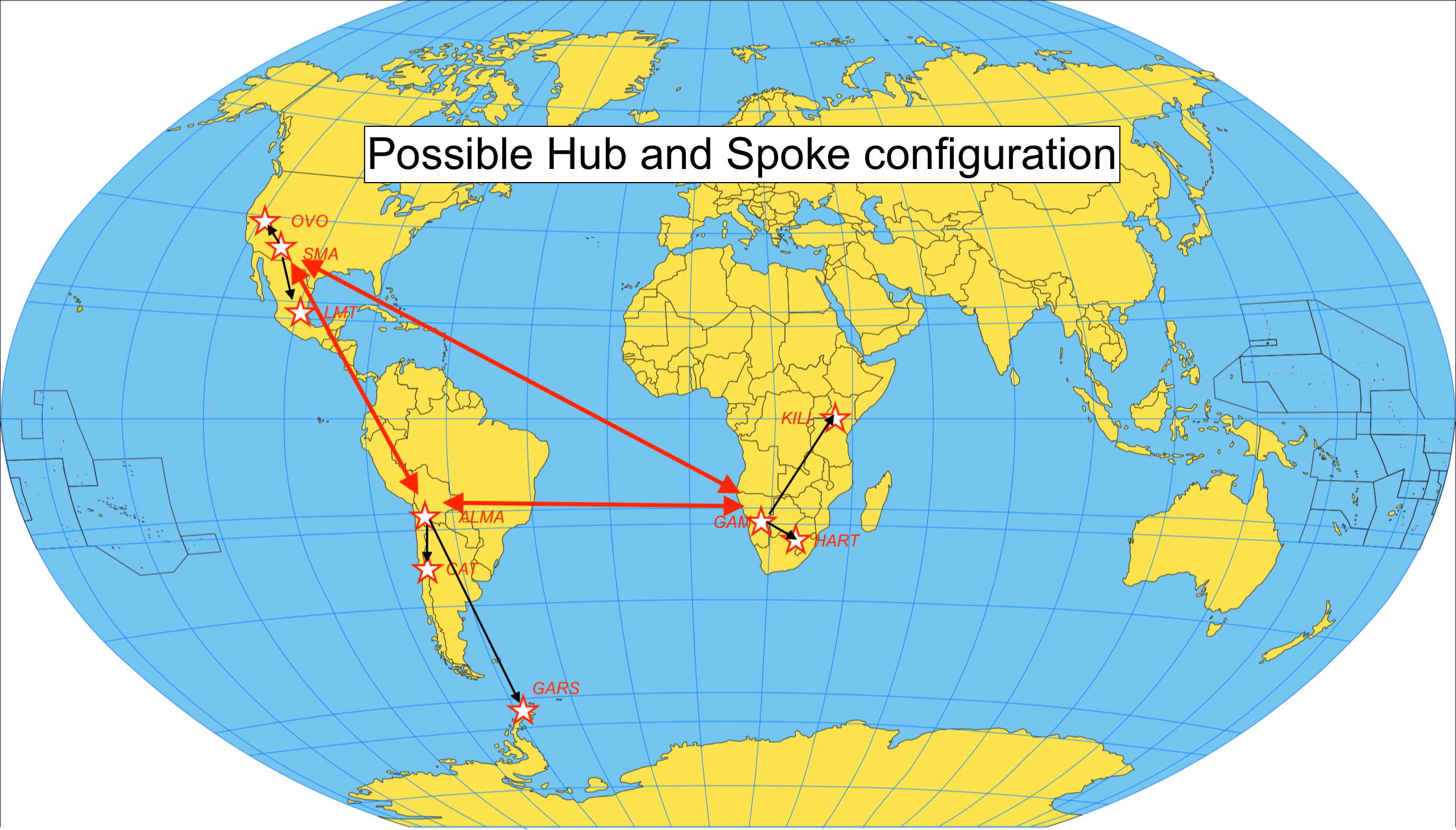}
    \caption{
    An example of sites that could form a hub-and-spoke configuration suitable for observing SgrA*, as used in our simulations. 
    The black lines connect smaller antennas to a hub that provides the calibration reference site. 
    These highly sensitive hub-based facets can then be connected together on their edges, or self-calibrated separately. Edited world map taken from Wikipedia. 
    \label{fig:wmap}}
\end{figure}

\begin{figure}
    \centering
    \includegraphics[width=0.6\textwidth]{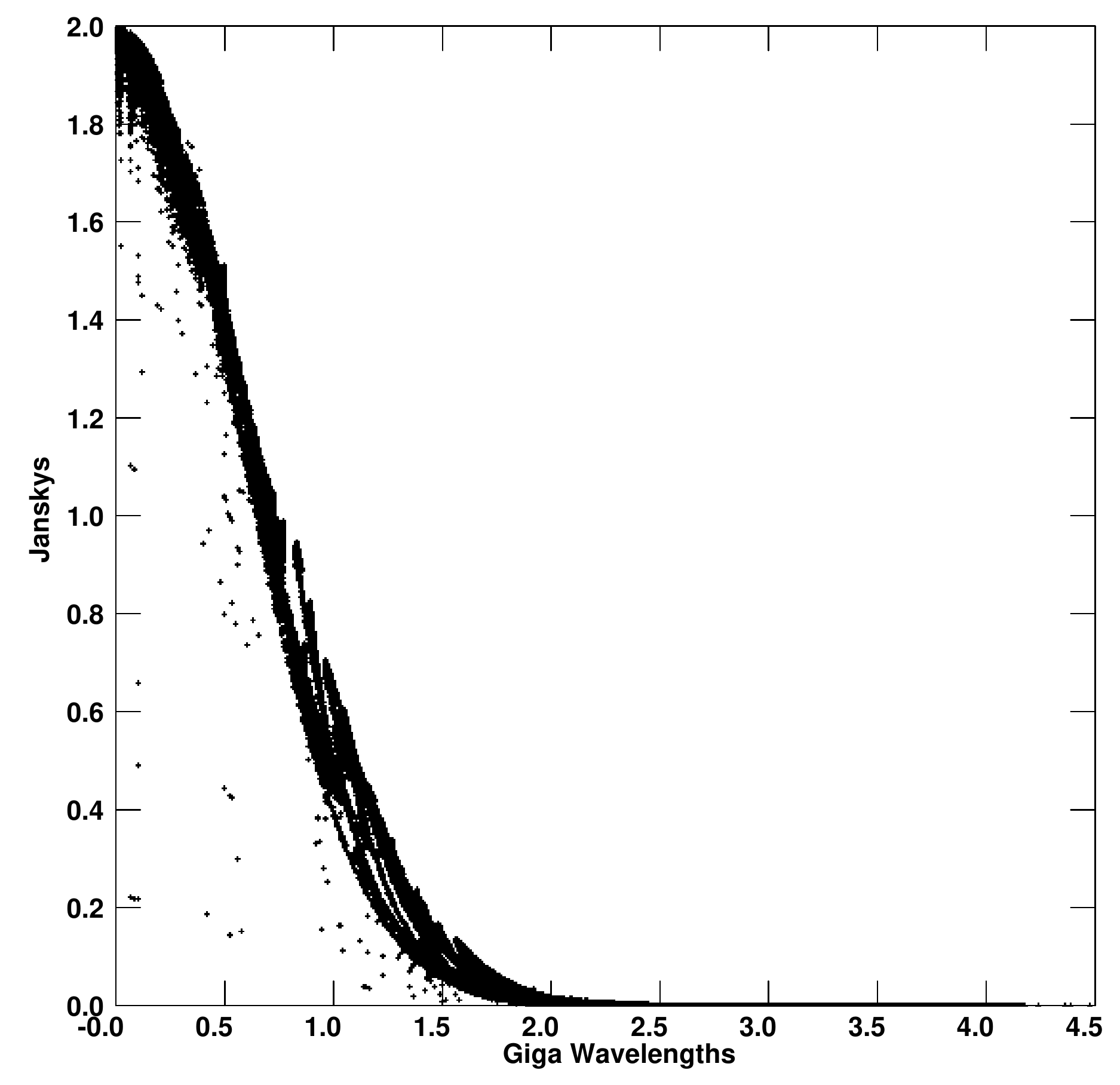}
    \caption{
    {Visibility amplitude for the simulations as a function of uv distance for observations of SgrA* at 115\,GHz.
    The model used in ARIS was a Gaussian based on the results of \citep{sgra_size}.
    showing that significant flux ($>$0.1\,Jy) is detectable out to 1.5G$\lambda$, corresponding to $\sim$4,000\,km. 
    This clearly shows why the baseline lengths need to be $<$1.5G$\lambda$, and a two step calibration (hub-to-spoke and facet-to-global solutions) is required. 
    \typeout{\bf some insights on the context on which this is relevant, to define the separation between various "short baselines sub-groups" in the hub and spoke configuration?}}
    \label{fig:sgra}}
\end{figure}

A successful application of frequency phase transfer techniques requires a direct detection, {within $\tau_{low}$ (related to the atmospheric coherence time)}, of the target source at the reference or lower frequency. 
In the presence of strong scatter-broadening the count-intuitive situation can arise where the source 
resolves at the lower frequency whilst at the higher frequency it does not, because the scattering is so much less.
This is the case for SgrA*, a main target for EHT and ngEHT studies, with an apparent size of 230$\times$140\,$\mu$as at 86\,GHz \citep{sgra_size}.
This challenges the FPT approach, but does not make it impossible. 
In essence what is required is that the array configuration consists of multiple highly sensitive hubs, surrounded by `spokes' to multiple smaller antennas that have shorter baselines, as shown in Figure \ref{fig:wmap}.
We thus did further ARIS simulations with Gaussian source models and a dense network of possible antenna sites.
These allow for the formation of hub-and-spoke facets that, in addition to being highly sensitive, then can be phase connected as the separation between the edges of the facets are less than the uv-range limit.
Given the size of the scattered source at 86\,GHz, this is still difficult; baselines longer than 2,250\,km would be hard to use in the calibration.
Therefore SgrA* is a strong driver for the requirement of the 85\,GHz band to cover 115-GHz, which would still be 
an integer ratio to the 230- and 345-GHz frequencies.
At 115-GHz the source size is 140$\times$115\,$\mu$as and baselines out to 1.5\,G$\lambda$ have correlated fluxes greater than 0.1\,Jy. Figure \ref{fig:sgra} shows the uv-amplitude plot of the simulated data with this model at  115\,GHz in the direction of SgrA*. 
In this case baselines out to $\sim$4,000\,km can be used in the calibration. Facets based in the Americas, Europe and Africa can then be connected.
Identification of suitable candidate hub sites in S. America, N. America, Africa and Europe should be undertaken; the main candidates are obvious and Figure \ref{fig:wmap} provides an example. 


\subsection{Paths to Astrometry at sub-mm wavelengths} \label{sec:mmastro}

A triple band receiver system and an array configuration with multiple antennas on a limited number of `hub' sites could enable various flavours of innovative astrometric measurements with the ngEHT array that would be impossible to achieve with single frequency observations, which we will discuss here.
The first flavour is `$\lambda$-Astrometry', 
which provides a precise bona-fide registration of the images at different frequencies. 
The second flavour is `conventional' or relative astrometry, which provides
a precise bona-fide measurement of the relative angular separation between a target and a reference source(s) at a given frequency.



\subsubsection{$\lambda$-astrometry}

We have estimated the astrometric accuracy at 255- or 340-GHz using the formulae for SFPR systematic astrometric error propagation in \citet{rioja_11a}, using 85-GHz as the reference frequency, with a source pair angular separation of 10$^o$, baseline lengths up to 6,000\,km and source switching times up to 10 minutes.
In this case the astrometric precision is predicted to be about 3\,$\mu$as and is dominated by the static ionospheric residuals and the large angular separation.

Source/Frequency Phase Referencing (SFPR) is a well demonstrated astrometric method and has been performed in the frequency range between 43- to 130-GHz \citep{rioja_15}.
The performance and ease of use was excellent, even though this was 
a three-fold increase over the previous frequency limit for astrometry.
Thus we expect it to work at ngEHT frequencies, i.e. from 85 to 340\,GHz, and SFPR should be straight forward with the current proposed array equipped with tri-band receivers.
On the other hand, the proposed dual-frequency receiver (230 and 340\,GHz) will struggle to provide $\lambda$-astrometry except in exceptional weather and for strong sources.
%



The scientific applications of $\lambda$-astrometry are extensively discussed in [Wu, J. in this journal] 
so further comments are not needed here. 
However we do point out that, once the frequency position offset is measured between 85\,GHz and either of the higher bands, one would be able to connect this offset to the ICRF if relative astrometry was performed at the lower frequency. 
This has been demonstrated at much lower frequencies in \citet{dodson_14}, for example. The application of MultiView to provide this is considered below.

\subsubsection{Relative Astrometry}


MultiView \citep{rioja_17} with simultaneous observations of multiple sources has the potential to provide a precise astrometry at 85\,GHz {using ngEHT. We note this is a regime beyond the scope of application of conventional phase referencing techniques.}
MultiView relies on the observations of multiple calibrators for a precise elimination of the dominant propagation {errors} from the observations, enabling ultra-precise relative astrometric measurements. 
It would provide bona-fide astrometry between sources, allowing registered dynamical measurements over time, to make movies of high impact individual sources or carry out statistical surveys of cosmological parallaxes.



MultiView has been demonstrated to provide precise corrections for the atmosphere at 1.4 \citep{rioja_17} and 8\,GHz \citep{hyland_22}, with on-going investigations at 43GHz. 
The calibrator solutions can be spatially interpolated to provide the atmospheric contributions at the target line of sight, which results in an effective angular separation equal to zero degrees. The residual systematic astrometric errors are estimated to be $\sim$\,1$\mu$as \citep[see the formulae in][]{rioja_20}.

One would require four (or more) antennas at each astrometric site to perform MultiView at 85\,GHz, to provide simultaneous observations of the target and three calibrator sources. 
ALMA would be able to perform this role when sub-arraying is implemented.

This configuration seems possible for ngEHT and would enable astrometric measurements as a stand alone instrument. Furthermore, the capability is definitely included in the ngVLA long baseline {array} planning \citep{ngvla_lbs}. 
A possible scenario is to perform combined measurements of the relative astrometry with the ngVLA at 85\,GHz with the $\lambda$-astrometry between 85 and 340\,GHz with the ngEHT; this would provide relative astrometry at 340\,GHz.

\subsection{Fundamental implications on network specifications}

{
Frequency phase transfer techniques result in a big boost of the sensitivity at 340\,GHz, compared to single-frequency observations, by increasing the coherence time. 
An increase in effective coherence time from 10 seconds to 60 min translates to a $\sim$20-fold decrease in the flux density threshold of detectable sources and a much larger source population within reach.
The FPT sensitivity is equivalent to single-frequency 340\,GHz observations with an array having 400-fold increase of bandwidth or a 4-fold increase in antenna diameter. 
}

\begin{figure}
    \centering
    \includegraphics[width=0.8\textwidth]{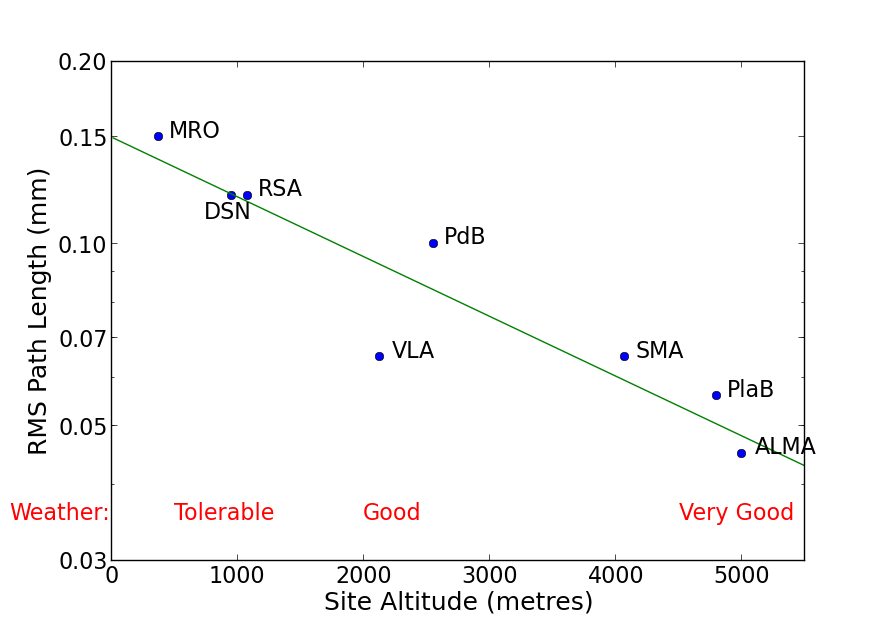}
    \caption{
    Implications of FPT benefits on antenna site selection. 
    The plot follows that of Figure 13.18 using the data in Table 13.4 in \citet{TMSv3}, showing the correlation between site altitude and empirical RMS pathlength measurements. 
    Overplotted, in red , is the correspondence  between altitudes and the simulation weather conditions explored in this paper, based on the RMS pathlength values from atmospheric models. 
    The use of FPT calibration relaxes the requirements for suitable sites for (sub)mm-VLBI observations and will allow Good sites, i.e. above 2,000m, of which there are many candidates.
    \label{fig:rms}}
\end{figure}

FPT techniques relax the constraints for suitable sites and/or weather conditions for (sub)mm-VLBI observations, with optimum outcomes having $\nu_{low}$ equal to 85-GHz.
This opens the possibilities for more ngEHT sites, which would provide better uv-coverage and better image fidelity.
Fewer constrains on the weather will provide extended windows of opportunity for observations, more observing time and improved temporal  monitoring.
%
%
Figure \ref{fig:rms} follows \citet[Figure 13.18 of][]{TMSv3}, showing
the empirical relationship between the altitude and the RMS phase fluctuations for a number of well-known telescope sites. 
Based on this and the RMS path-length values for the weather models used in our simulations 
we overplot the estimated altitudes corresponding to V, G and T weather conditions.
For the best performance with 85\,GHz as the reference frequency, site altitudes greater than 2,000m are acceptable, based on the discussions in Section \ref{sec:85}. 
For comparison, with 230\,GHz as the reference frequency, site altitudes greater than 4,000m are desired, based on the discussions in Section \ref{sec:230}.

\typeout{Space is an antenna location site}
Another interesting outcome from the frequency phase transfer-group of
calibration methods is that they enable space VLBI at ngEHT frequencies, for highest angular resolution imaging with enhanced sensitivity and ultra precise astrometry.
Orbit errors, like tropospheric errors, are non-dispersive hence FPT {and SFPR} methods will resolve these
\citep{rioja_11b}.

\subsection{Requirements \& Guidelines} 
\label{sec:req}


{
The requirements for the application of frequency phase transfer techniques have been discussed at length in Section 4 of \citep[][]{rioja_11a}. Here we summarise the points particularly relevant to ngEHT.}


\begin{itemize}

\item Simultaneous observations at the multiple frequencies is an absolute key requirement, as this is the only way to fully sample the fast tropospheric fluctuations at ngEHT frequencies.

\item Direct detections at the reference frequency within the relatively short calibration timescale, $\tau_{low}$, is a fundamental requirement; the upper bound for $\tau_{low}$ sits between the atmospheric coherence time of the two frequencies ($\nu_{low}$ and $\nu_{high}$).
Note that, since phase errors are also scaled, it is important to optimise the SNR of the dectections. Having 85-GHz as the reference frequency alleviates this requirement with the lower system noise and longer $\tau_{low}$ values, and in-general larger source flux densities, compared to 230-GHz. 

\item We strongly recommended that the $\nu_{low}$ and $\nu_{high}$ frequencies have an integer ratio to avoid phase-ambiguity-related problems in the analysis. 
Note that it is sufficient that the selected frequency reference points are within or close to the recorded frequency bands. 
Non-integer frequency ratios will, in general, introduce phase offsets and jumps which would have to be addressed separately.

\item SgrA* requires observations at 115\,GHz and a hub-and-spoke array configuration, so that all stations have sufficiently short sensitive baselines for detections.

\item We recommend a tri-band receiver system with 85-, 230- and 340-GHz bands. The recorded bandwidth of the 85-GHz system should cover 115\,GHz. Care should be take to ensure that integer frequency ratios are possible within the recorded bands.

\item Instrumental effects should be less than the residual atmospheric effects post frequency phase transfer analysis. %
For optimum performance it is required that instrumental effects, including between bands, must be measurable and stable to be better than the equivalent of 1$\mu$as or 5 picoseconds. 
Larger residual instrumental effects might result in these being the dominant source of errors and limit the performance, where not thermally limited. 

\item MultiView will require a small number of `hub' sites, able to carry out simultaneous observations along different lines of sight. This would be sub-arraying for ALMA and for multiple-antenna sites the requirement is for four antennas. 
\end{itemize}

\section{Summary and conclusions}\label{sec:conc}

Frequency phase transfer techniques hold the potential to revolutionise (sub)mm-VLBI observations with ngEHT, making possible observations at 340\,GHz where single-frequency observations are impossible. FPT and SFPR enable  an adaptive phase-based calibration of the tropospheric fluctuations at the target frequency, using simultaneous observations at a reference frequency, of a common source.
The best outcomes are expected from the proposed tri-band receiver, which encompass the addition of a 85-GHz band  to the original system design of a dual-frequency receiver covering the 230 and 340-GHz bands. Our conclusions are based on the explorations of comparative ngEHT performance at 340 GHz presented in this paper.

Our metrics are primarily the quality of the FPT calibration at 340\,GHz, testing the improvements of phase stability and coherence, and the lengths of time over which this is sustained. 
This FPT-coherence is orders of magnitude longer than the atmospheric-coherence time at the higher ngEHT frequencies. 
This results in reduced flux-detection limits and increased dynamic range and effective angular resolution in the images. 

We show that using 85\,GHz as reference in frequency phase transfer methods provides for robust operations of the ngEHT array at 340\,GHz under a wider range of weather conditions and from sites that would be marginal for single-frequency 340-GHz VLBI observations.
This means that observations at the highest frequencies with ngEHT will be routinely viable over the whole year, for source monitoring and movies of interesting sources.
Success using 230\,GHz as the reference frequency is possible, but the constraints are more severe.
That is, only under Very Good weather conditions ($C_w$ equal to 0.5, or site altitudes above $\sim$4,000m) would 230\,GHz be a reliable reference frequency, whereas 85\,GHz would be reliable even under Good weather conditions  ($C_w$ equal to 1, or site altitudes above $\sim$2,000m).

A fundamental requirement is for direct detections of the target source at the reference frequency within the $\tau_{low}$ timescale.
This poses a more severe constraint at 230\,GHz, compared to 85\,GHz. 
Based on the ALMA calibrator list, 20\% of sources are detectable by ngEHT at 230-GHz whereas 90\% are detectable at 85-GHz.

Including the 85-GHz band allows for a richer range of astrometric observations. 
SFPR provides for $\mu$-as bona-fide astrometry, for cross-band image registration ($\lambda$-astrometry) between the three bands, 
and potentially for relative astrometry up to 340\,GHz, where not thermally limited.  
The latter requires four antennas co-observing at a subgroup of sites to perform MultiView astrometry at 85\,GHz with $\mu$-as precision.

For observing highly scatter-broaden sources such as SgrA* with frequency phase transfer methods we recommend a hub-and-spoke configuration for the array, and reference observations at a frequency of 115\,GHz.

The benefits of frequency phase transfer techniques extend to the (sub)mm space VLBI regime, 
where the increase in spatial resolution by orders of magnitude and capacity for ultra precise astrometry will be a game changer.  

Finally, a stable and well calibrated multi-frequency instrument is required, 
with control systems in place to reduce the level of instrumental errors to less than the residuals from the frequency phase transfer methods.
We provide specifications such that they do not become the dominant source of errors and limit the performance.

Based on all of the above we conclude that the tri-band system will have a critical impact on the ngEHT scientific outcomes.
Frequency phase transfer techniques will enable for (sub)mm VLBI  the full range of functionalities available for the cm-VLBI regime, with $\mu$-as resolution. That is:
high sensitivity, high fidelity images, ultra-precise astrometric measurements, routinely-viable year-round imaging and using ground or space VLBI observations.
These provide the basis for carrying out monitoring, making movies and performing supermassive black-hole demographic studies, 
which are key scientific drivers for ngEHT. 

Adaptive phase-based FPT and SFPR methods overcome the challenge introduced by fast tropospheric fluctuations and the severe  limitations imposed by them at ngEHT frequencies.
Its application will provide a
transformational impact in the ngEHT array capabilities, changing the  (sub)mm-VLBI world. 
Therefore, we strongly recommend that establishing this capability should be an area of highest priority for  the ngEHT.


\begin{thebibliography}{999}

\bibitem[{Event Horizon Telescope Collaboration}(2019)]{eht_bh}
{Event Horizon Telescope Collaboration}.
\newblock {First M87 Event Horizon Telescope Results. I. The Shadow of the
  Supermassive Black Hole}.
\newblock {\em \apjl} {\bf 2019}, {\em 875},~L1,
  \href{http://xxx.lanl.gov/abs/1906.11238}{{\normalfont
  [arXiv:astro-ph.GA/1906.11238]}}.
\newblock {\url{https://doi.org/10.3847/2041-8213/ab0ec7}}.

\bibitem[{Event Horizon Telescope Collaboration} \em{et~al.}(2022){Event
  Horizon Telescope Collaboration}, {Akiyama}, {Alberdi}, {Alef}, {Algaba},
  {Anantua}, {Asada}, {Azulay}, {Bach}, {Baczko}, {Ball}, {Balokovi{\'c}},
  {Barrett}, {Baub{\"o}ck}, {Benson}, {Bintley}, {Blackburn}, {Blundell},
  {Bouman}, {Bower}, {Boyce}, {Bremer}, {Brinkerink}, {Brissenden}, {Britzen},
  {Broderick}, {Broguiere}, {Bronzwaer}, {Bustamante}, {Byun}, {Carlstrom},
  {Ceccobello}, {Chael}, {Chan}, {Chatterjee}, {Chatterjee}, {Chen}, {Chen},
  {Cheng}, {Cho}, {Christian}, {Conroy}, {Conway}, {Cordes}, {Crawford},
  {Crew}, {Cruz-Osorio}, {Cui}, {Davelaar}, {Laurentis}, {Deane}, {Dempsey},
  {Desvignes}, {Dexter}, {Dhruv}, {Doeleman}, {Dougal}, {Dzib}, {Eatough},
  {Emami}, {Falcke}, {Farah}, {Fish}, {Fomalont}, {Ford}, {Fraga-Encinas},
  {Freeman}, {Friberg}, {Fromm}, {Fuentes}, {Galison}, {Gammie}, {Garc{\'\i}a},
  {Gentaz}, {Georgiev}, {Goddi}, {Gold}, {G{\'o}mez-Ruiz}, {G{\'o}mez}, {Gu},
  {Gurwell}, {Hada}, {Haggard}, {Haworth}, {Hecht}, {Hesper}, {Heumann}, {Ho},
  {Ho}, {Honma}, {Huang}, {Huang}, {Hughes}, {Ikeda}, {Impellizzeri}, {Inoue},
  {Issaoun}, {James}, {Jannuzi}, {Janssen}, {Jeter}, {Jiang},
  {Jim{\'e}nez-Rosales}, {Johnson}, {Jorstad}, {Joshi}, {Jung}, {Karami},
  {Karuppusamy}, {Kawashima}, {Keating}, {Kettenis}, {Kim}, {Kim}, {Kim},
  {Kim}, {Kino}, {Koay}, {Kocherlakota}, {Kofuji}, {Koch}, {Koyama}, {Kramer},
  {Kramer}, {Krichbaum}, {Kuo}, {Bella}, {Lauer}, {Lee}, {Lee}, {Leung},
  {Levis}, {Li}, {Lico}, {Lindahl}, {Lindqvist}, {Lisakov}, {Liu}, {Liu},
  {Liuzzo}, {Lo}, {Lobanov}, {Loinard}, {Lonsdale}, {Lu}, {Mao}, {Marchili},
  {Markoff}, {Marrone}, {Marscher}, {Mart{\'\i}-Vidal}, {Matsushita},
  {Matthews}, {Medeiros}, {Menten}, {Michalik}, {Mizuno}, {Mizuno}, {Moran},
  {Moriyama}, {Moscibrodzka}, {M{\"u}ller}, {Mus}, {Musoke}, {Myserlis},
  {Nadolski}, {Nagai}, {Nagar}, {Nakamura}, {Narayan}, {Narayanan},
  {Natarajan}, {Nathanail}, {Fuentes}, {Neilsen}, {Neri}, {Ni}, {Noutsos},
  {Nowak}, {Oh}, {Okino}, {Olivares}, {Ortiz-Le{\'o}n}, {Oyama}, {{\"O}zel},
  {Palumbo}, {Paraschos}, {Park}, {Parsons}, {Patel}, {Pen}, {Pesce},
  {Pi{\'e}tu}, {Plambeck}, {PopStefanija}, {Porth}, {P{\"o}tzl}, {Prather},
  {Preciado-L{\'o}pez}, {Psaltis}, {Pu}, {Ramakrishnan}, {Rao}, {Rawlings},
  {Raymond}, {Rezzolla}, {Ricarte}, {Ripperda}, {Roelofs}, {Rogers}, {Ros},
  {Romero-Ca{\~n}izales}, {Roshanineshat}, {Rottmann}, {Roy}, {Ruiz},
  {Ruszczyk}, {Rygl}, {S{\'a}nchez}, {S{\'a}nchez-Arg{\"u}elles},
  {S{\'a}nchez-Portal}, {Sasada}, {Satapathy}, {Savolainen}, {Schloerb},
  {Schonfeld}, {Schuster}, {Shao}, {Shen}, {Small}, {Sohn}, {SooHoo},
  {Souccar}, {Sun}, {Tazaki}, {Tetarenko}, {Tiede}, {Tilanus}, {Titus},
  {Torne}, {Traianou}, {Trent}, {Trippe}, {Turk}, {van Bemmel}, {van
  Langevelde}, {van Rossum}, {Vos}, {Wagner}, {Ward-Thompson}, {Wardle},
  {Weintroub}, {Wex}, {Wharton}, {Wielgus}, {Wiik}, {Witzel}, {Wondrak},
  {Wong}, {Wu}, {Yamaguchi}, {Yoon}, {Young}, {Young}, {Younsi}, {Yuan},
  {Yuan}, {Zensus}, {Zhang}, {Zhao}, {Zhao}, {Agurto}, {Allardi}, {Amestica},
  {Araneda}, {Arriagada}, {Berghuis}, {Bertarini}, {Berthold}, {Blanchard},
  {Brown}, {C{\'a}rdenas}, {Cantzler}, {Caro}, {Castillo-Dom{\'\i}nguez},
  {Chan}, {Chang}, {Chang}, {Chang}, {Chang}, {Chen}, {Chilson}, {Chuter},
  {Ciechanowicz}, {Colin-Beltran}, {Coulson}, {Crowley}, {Degenaar},
  {Dornbusch}, {Dur{\'a}n}, {Everett}, {Faber}, {Forster}, {Fuchs}, {Gale},
  {Geertsema}, {Gonz{\'a}lez}, {Graham}, {Gueth}, {Halverson}, {Han}, {Han},
  {Hasegawa}, {Hern{\'a}ndez-Rebollar}, {Herrera}, {Herrero-Illana},
  {Heyminck}, {Hirota}, {Hoge}, {Hostler Schimpf}, {Howie}, {Huang}, {Jiang},
  {Jinchi}, {John}, {Kimura}, {Klein}, {Kubo}, {Kuroda}, {Kwon}, {Lacasse},
  {Laing}, {Leitch}, {Li}, {Liu}, {Liu}, {Lin}, {Lu}, {Mac-Auliffe},
  {Martin-Cocher}, {Matulonis}, {Maute}, {Messias}, {Meyer-Zhao},
  {Monta{\~n}a}, {Montenegro-Montes}, {Montgomerie}, {Moreno Nolasco},
  {Muders}, {Nishioka}, {Norton}, {Nystrom}, {Ogawa}, {Olivares}, {Oshiro},
  {P{\'e}rez-Beaupuits}, {Parra}, {Phillips}, {Poirier}, {Pradel}, {Qiu},
  {Raffin}, {Rahlin}, {Ram{\'\i}rez}, {Ressler}, {Reynolds},
  {Rodr{\'\i}guez-Montoya}, {Saez-Madain}, {Santana}, {Shaw}, {Shirkey},
  {Silva}, {Snow}, {Sousa}, {Sridharan}, {Stahm}, {Stark}, {Test},
  {Torstensson}, {Venegas}, {Walther}, {Wei}, {White}, {Wieching}, {Wijnands},
  {Wouterloot}, {Yu}, {Yu (于威)}, and {Zeballos}]{eht_sgra}
{Event Horizon Telescope Collaboration}.; {Akiyama}, K.; {Alberdi}, A.; {Alef},
  W.; {Algaba}, J.C.; {Anantua}, R.; {Asada}, K.; {Azulay}, R.; {Bach}, U.;
  {Baczko}, A.K.;  et~al.
\newblock {First Sagittarius A* Event Horizon Telescope Results. I. The Shadow
  of the Supermassive Black Hole in the Center of the Milky Way}.
\newblock {\em \apjl} {\bf 2022}, {\em 930},~L12.
\newblock {\url{https://doi.org/10.3847/2041-8213/ac6674}}.

\bibitem[{Event Horizon Telescope Collaboration} \em{et~al.}(2019){Event
  Horizon Telescope Collaboration}, {Akiyama}, {Alberdi}, {Alef}, {Asada},
  {Azulay}, {Baczko}, {Ball}, {Balokovi{\'c}}, {Barrett}, {Bintley},
  {Blackburn}, {Boland}, {Bouman}, {Bower}, {Bremer}, {Brinkerink},
  {Brissenden}, {Britzen}, {Broderick}, {Broguiere}, {Bronzwaer}, {Byun},
  {Carlstrom}, {Chael}, {Chan}, {Chatterjee}, {Chatterjee}, {Chen}, {Chen},
  {Cho}, {Christian}, {Conway}, {Cordes}, {Crew}, {Cui}, {Davelaar}, {De
  Laurentis}, {Deane}, {Dempsey}, {Desvignes}, {Dexter}, {Doeleman}, {Eatough},
  {Falcke}, {Fish}, {Fomalont}, {Fraga-Encinas}, {Freeman}, {Friberg}, {Fromm},
  {G{\'o}mez}, {Galison}, {Gammie}, {Garc{\'\i}a}, {Gentaz}, {Georgiev},
  {Goddi}, {Gold}, {Gu}, {Gurwell}, {Hada}, {Hecht}, {Hesper}, {Ho}, {Ho},
  {Honma}, {Huang}, {Huang}, {Hughes}, {Ikeda}, {Inoue}, {Issaoun}, {James},
  {Jannuzi}, {Janssen}, {Jeter}, {Jiang}, {Johnson}, {Jorstad}, {Jung},
  {Karami}, {Karuppusamy}, {Kawashima}, {Keating}, {Kettenis}, {Kim}, {Kim},
  {Kim}, {Kino}, {Koay}, {Koch}, {Koyama}, {Kramer}, {Kramer}, {Krichbaum},
  {Kuo}, {Lauer}, {Lee}, {Li}, {Li}, {Lindqvist}, {Liu}, {Liuzzo}, {Lo},
  {Lobanov}, {Loinard}, {Lonsdale}, {Lu}, {MacDonald}, {Mao}, {Markoff},
  {Marrone}, {Marscher}, {Mart{\'\i}-Vidal}, {Matsushita}, {Matthews},
  {Medeiros}, {Menten}, {Mizuno}, {Mizuno}, {Moran}, {Moriyama},
  {Moscibrodzka}, {M{\"u}ller}, {Nagai}, {Nagar}, {Nakamura}, {Narayan},
  {Narayanan}, {Natarajan}, {Neri}, {Ni}, {Noutsos}, {Okino}, {Olivares},
  {Oyama}, {{\"O}zel}, {Palumbo}, {Patel}, {Pen}, {Pesce}, {Pi{\'e}tu},
  {Plambeck}, {PopStefanija}, {Porth}, {Prather}, {Preciado-L{\'o}pez},
  {Psaltis}, {Pu}, {Ramakrishnan}, {Rao}, {Rawlings}, {Raymond}, {Rezzolla},
  {Ripperda}, {Roelofs}, {Rogers}, {Ros}, {Rose}, {Roshanineshat}, {Rottmann},
  {Roy}, {Ruszczyk}, {Ryan}, {Rygl}, {S{\'a}nchez}, {S{\'a}nchez-Arguelles},
  {Sasada}, {Savolainen}, {Schloerb}, {Schuster}, {Shao}, {Shen}, {Small},
  {Sohn}, {SooHoo}, {Tazaki}, {Tiede}, {Tilanus}, {Titus}, {Toma}, {Torne},
  {Trent}, {Trippe}, {Tsuda}, {van Bemmel}, {van Langevelde}, {van Rossum},
  {Wagner}, {Wardle}, {Weintroub}, {Wex}, {Wharton}, {Wielgus}, {Wong}, {Wu},
  {Young}, {Young}, {Younsi}, {Yuan}, {Yuan}, {Zensus}, {Zhao}, {Zhao}, {Zhu},
  {Farah}, {Meyer-Zhao}, {Michalik}, {Nadolski}, {Nishioka}, {Pradel},
  {Primiani}, {Souccar}, {Vertatschitsch}, and {Yamaguchi}]{eht_m87_imaging}
{Event Horizon Telescope Collaboration}.; {Akiyama}, K.; {Alberdi}, A.; {Alef},
  W.; {Asada}, K.; {Azulay}, R.; {Baczko}, A.K.; {Ball}, D.; {Balokovi{\'c}},
  M.; {Barrett}, J.;  et~al.
\newblock {First M87 Event Horizon Telescope Results. IV. Imaging the Central
  Supermassive Black Hole}.
\newblock {\em \apjl} {\bf 2019}, {\em 875},~L4,
  \href{http://xxx.lanl.gov/abs/1906.11241}{{\normalfont
  [arXiv:astro-ph.GA/1906.11241]}}.
\newblock {\url{https://doi.org/10.3847/2041-8213/ab0e85}}.

\bibitem[{Asaki} \em{et~al.}(1996){Asaki}, {Saito}, {Kawabe}, {Morita}, and
  {Sasao}]{asaki_fpt_96}
{Asaki}, Y.; {Saito}, M.; {Kawabe}, R.; {Morita}, K.I.; {Sasao}, T.
\newblock {Phase compensation experiments with the paired antennas method}.
\newblock {\em Radio Science} {\bf 1996}, {\em 31},~1615--1626.
\newblock {\url{https://doi.org/10.1029/96RS02629}}.

\bibitem[{Carilli} and {Holdaway}(1999)]{carilli_99}
{Carilli}, C.L.; {Holdaway}, M.A.
\newblock {Tropospheric phase calibration in millimeter interferometry}.
\newblock {\em Radio Science} {\bf 1999}, {\em 34},~817--840,
  \href{http://xxx.lanl.gov/abs/astro-ph/9904248}{{\normalfont
  [astro-ph/9904248]}}.
\newblock {\url{https://doi.org/10.1029/1999RS900048}}.

\bibitem[{Middelberg} \em{et~al.}(2005){Middelberg}, {Roy}, {Walker}, and
  {Falcke}]{middelberg_05}
{Middelberg}, E.; {Roy}, A.L.; {Walker}, R.C.; {Falcke}, H.
\newblock {VLBI observations of weak sources using fast frequency switching}.
\newblock {\em \aap} {\bf 2005}, {\em 433},~897--909,
  \href{http://xxx.lanl.gov/abs/astro-ph/0412564}{{\normalfont
  [astro-ph/0412564]}}.
\newblock {\url{https://doi.org/10.1051/0004-6361:20042078}}.

\bibitem[{Rioja} and {Dodson}(2011)]{rioja_11a}
{Rioja}, M.; {Dodson}, R.
\newblock {High-precision Astrometric Millimeter Very Long Baseline
  Interferometry Using a New Method for Atmospheric Calibration}.
\newblock {\em AJ} {\bf 2011}, {\em 141},~114,
  \href{http://xxx.lanl.gov/abs/arXiv:1101.2051}{{\normalfont
  [arXiv:astro-ph.IM/arXiv:1101.2051]}}.
\newblock {\url{https://doi.org/10.1088/0004-6256/141/4/114}}.

\bibitem[{Rioja} \em{et~al.}(2015){Rioja}, {Dodson}, {Jung}, and
  {Sohn}]{rioja_15}
{Rioja}, M.J.; {Dodson}, R.; {Jung}, T.; {Sohn}, B.W.
\newblock {The Power of Simultaneous Multi-Frequency Observations for mm-VLBI:
  Astrometry up to 130 GHz with the KVN}.
\newblock {\em AJ} {\bf 2015}, {\em 150},~202,
  \href{http://xxx.lanl.gov/abs/1509.02621}{{\normalfont
  [arXiv:astro-ph.IM/1509.02621]}}.
\newblock {\url{https://doi.org/10.1088/0004-6256/150/6/202}}.

\bibitem[{Rioja} and {Dodson}(2020)]{rioja_20}
{Rioja}, M.J.; {Dodson}, R.
\newblock {Precise radio astrometry and new developments for the
  next-generation of instruments}.
\newblock {\em \aapr} {\bf 2020}, {\em 28},~6,
  \href{http://xxx.lanl.gov/abs/2010.02156}{{\normalfont
  [arXiv:astro-ph.IM/2010.02156]}}.
\newblock {\url{https://doi.org/10.1007/s00159-020-00126-z}}.

\bibitem[{Dodson} \em{et~al.}(2014){Dodson}, {Rioja}, {Jung}, {Sohn}, {Byun},
  {Cho}, {Lee}, {Kim}, {Kim}, {Oh}, {Han}, {Je}, {Chung}, {Wi}, {Kang}, {Lee},
  {Chung}, {Kim}, {Kim}, {Lee}, {Roh}, {Oh}, {Yeom}, {Song}, and
  {Kang}]{dodson_14}
{Dodson}, R.; {Rioja}, M.J.; {Jung}, T.H.; {Sohn}, B.W.; {Byun}, D.Y.; {Cho},
  S.H.; {Lee}, S.S.; {Kim}, J.; {Kim}, K.T.; {Oh}, C.S.;  et~al.
\newblock {Astrometrically Registered Simultaneous Observations of the 22 GHz
  H$_{2}$O and 43 GHz SiO Masers toward R Leonis Minoris Using KVN and
  Source/Frequency Phase Referencing}.
\newblock {\em AJ} {\bf 2014}, {\em 148},~97,
  \href{http://xxx.lanl.gov/abs/arXiv:1408.3513}{{\normalfont
  [arXiv:astro-ph.IM/arXiv:1408.3513]}}.
\newblock {\url{https://doi.org/10.1088/0004-6256/148/5/97}}.

\bibitem[{Dodson} \em{et~al.}(2017){Dodson}, {Rioja}, {Molina}, and
  {G{\'o}mez}]{dodson_17}
{Dodson}, R.; {Rioja}, M.J.; {Molina}, S.N.; {G{\'o}mez}, J.L.
\newblock {High-precision Astrometric Millimeter Very Long Baseline
  Interferometry Using a New Method for Multi-frequency Calibration}.
\newblock {\em \apj} {\bf 2017}, {\em 834},~177,
  \href{http://xxx.lanl.gov/abs/1612.02958}{{\normalfont
  [arXiv:astro-ph.IM/1612.02958]}}.
\newblock {\url{https://doi.org/10.3847/1538-4357/834/2/177}}.

\bibitem[{Zhao} \em{et~al.}(2018){Zhao}, {Algaba}, {Lee}, {Jung}, {Dodson},
  {Rioja}, {Byun}, {Hodgson}, {Kang}, {Kim}, {Kim}, {Kim}, {Kim}, {Kino},
  {Miyazaki}, {Park}, {Trippe}, and {Wajima}]{zhao_17}
{Zhao}, G.Y.; {Algaba}, J.C.; {Lee}, S.S.; {Jung}, T.; {Dodson}, R.; {Rioja},
  M.; {Byun}, D.Y.; {Hodgson}, J.; {Kang}, S.; {Kim}, D.W.;  et~al.
\newblock {The Power of Simultaneous Multi-frequency Observations for mm-VLBI:
  Beyond Frequency Phase Transfer}.
\newblock {\em \aj} {\bf 2018}, {\em 155},~26,
  \href{http://xxx.lanl.gov/abs/1712.06243}{{\normalfont
  [arXiv:astro-ph.IM/1712.06243]}}.
\newblock {\url{https://doi.org/10.3847/1538-3881/aa99e0}}.

\bibitem[{Rioja} \em{et~al.}(2011){Rioja}, {Dodson}, {Malarecki}, and
  {Asaki}]{rioja_11b}
{Rioja}, M.; {Dodson}, R.; {Malarecki}, J.; {Asaki}, Y.
\newblock {Exploration of Source Frequency Phase Referencing Techniques for
  Astrometry and Observations of Weak Sources with High Frequency Space Very
  Long Baseline Interferometry}.
\newblock {\em AJ} {\bf 2011}, {\em 142},~157,
  \href{http://xxx.lanl.gov/abs/arXiv:1110.0267}{{\normalfont
  [arXiv:astro-ph.IM/arXiv:1110.0267]}}.
\newblock {\url{https://doi.org/10.1088/0004-6256/142/5/157}}.

\bibitem[{Asaki} \em{et~al.}(2007){Asaki}, {Sudou}, {Kono}, {Doi}, {Dodson},
  {Pradel}, {Murata}, {Mochizuki}, {Edwards}, {Sasao}, and {Fomalont}]{a07}
{Asaki}, Y.; {Sudou}, H.; {Kono}, Y.; {Doi}, A.; {Dodson}, R.; {Pradel}, N.;
  {Murata}, Y.; {Mochizuki}, N.; {Edwards}, P.G.; {Sasao}, T.;  et~al.
\newblock {Verification of the Effectiveness of VSOP-2 Phase Referencing with a
  Newly Developed Simulation Tool, ARIS}.
\newblock {\em \pasj} {\bf 2007}, {\em 59},~397--418,
  \href{http://xxx.lanl.gov/abs/0707.0558}{{\normalfont [0707.0558]}}.
\newblock {\url{https://doi.org/10.1093/pasj/59.2.397}}.

\bibitem[{Event Horizon Telescope Collaboration} \em{et~al.}(2019){Event
  Horizon Telescope Collaboration}, {Akiyama}, {Alberdi}, {Alef}, {Asada},
  {Azulay}, {Baczko}, {Ball}, {Balokovi{\'c}}, {Barrett}, {Bintley},
  {Blackburn}, {Boland}, {Bouman}, {Bower}, {Bremer}, {Brinkerink},
  {Brissenden}, {Britzen}, {Broderick}, {Broguiere}, {Bronzwaer}, {Byun},
  {Carlstrom}, {Chael}, {Chan}, {Chatterjee}, {Chatterjee}, {Chen}, {Chen},
  {Cho}, {Christian}, {Conway}, {Cordes}, {Crew}, {Cui}, {Davelaar}, {De
  Laurentis}, {Deane}, {Dempsey}, {Desvignes}, {Dexter}, {Doeleman}, {Eatough},
  {Falcke}, {Fish}, {Fomalont}, {Fraga-Encinas}, {Friberg}, {Fromm},
  {G{\'o}mez}, {Galison}, {Gammie}, {Garc{\'\i}a}, {Gentaz}, {Georgiev},
  {Goddi}, {Gold}, {Gu}, {Gurwell}, {Hada}, {Hecht}, {Hesper}, {Ho}, {Ho},
  {Honma}, {Huang}, {Huang}, {Hughes}, {Ikeda}, {Inoue}, {Issaoun}, {James},
  {Jannuzi}, {Janssen}, {Jeter}, {Jiang}, {Johnson}, {Jorstad}, {Jung},
  {Karami}, {Karuppusamy}, {Kawashima}, {Keating}, {Kettenis}, {Kim}, {Kim},
  {Kim}, {Kino}, {Koay}, {Koch}, {Koyama}, {Kramer}, {Kramer}, {Krichbaum},
  {Kuo}, {Lauer}, {Lee}, {Li}, {Li}, {Lindqvist}, {Liu}, {Liuzzo}, {Lo},
  {Lobanov}, {Loinard}, {Lonsdale}, {Lu}, {MacDonald}, {Mao}, {Markoff},
  {Marrone}, {Marscher}, {Mart{\'\i}-Vidal}, {Matsushita}, {Matthews},
  {Medeiros}, {Menten}, {Mizuno}, {Mizuno}, {Moran}, {Moriyama},
  {Moscibrodzka}, {M{\"u}ller}, {Nagai}, {Nagar}, {Nakamura}, {Narayan},
  {Narayanan}, {Natarajan}, {Neri}, {Ni}, {Noutsos}, {Okino}, {Olivares},
  {Ortiz-Le{\'o}n}, {Oyama}, {{\"O}zel}, {Palumbo}, {Patel}, {Pen}, {Pesce},
  {Pi{\'e}tu}, {Plambeck}, {PopStefanija}, {Porth}, {Prather},
  {Preciado-L{\'o}pez}, {Psaltis}, {Pu}, {Ramakrishnan}, {Rao}, {Rawlings},
  {Raymond}, {Rezzolla}, {Ripperda}, {Roelofs}, {Rogers}, {Ros}, {Rose},
  {Roshanineshat}, {Rottmann}, {Roy}, {Ruszczyk}, {Ryan}, {Rygl},
  {S{\'a}nchez}, {S{\'a}nchez-Arguelles}, {Sasada}, {Savolainen}, {Schloerb},
  {Schuster}, {Shao}, {Shen}, {Small}, {Sohn}, {SooHoo}, {Tazaki}, {Tiede},
  {Tilanus}, {Titus}, {Toma}, {Torne}, {Trent}, {Trippe}, {Tsuda}, {van
  Bemmel}, {van Langevelde}, {van Rossum}, {Wagner}, {Wardle}, {Weintroub},
  {Wex}, {Wharton}, {Wielgus}, {Wong}, {Wu}, {Young}, {Young}, {Younsi},
  {Yuan}, {Yuan}, {Zensus}, {Zhao}, {Zhao}, {Zhu}, {Algaba}, {Allardi},
  {Amestica}, {Bach}, {Beaudoin}, {Benson}, {Berthold}, {Blanchard},
  {Blundell}, {Bustamente}, {Cappallo}, {Castillo-Dom{\'\i}nguez}, {Chang},
  {Chang}, {Chang}, {Chen}, {Chilson}, {Chuter}, {C{\'o}rdova Rosado},
  {Coulson}, {Crawford}, {Crowley}, {David}, {Derome}, {Dexter}, {Dornbusch},
  {Dudevoir}, {Dzib}, {Eckert}, {Erickson}, {Everett}, {Faber}, {Farah},
  {Fath}, {Folkers}, {Forbes}, {Freund}, {G{\'o}mez-Ruiz}, {Gale}, {Gao},
  {Geertsema}, {Graham}, {Greer}, {Grosslein}, {Gueth}, {Halverson}, {Han},
  {Han}, {Hao}, {Hasegawa}, {Henning}, {Hern{\'a}ndez-G{\'o}mez},
  {Herrero-Illana}, {Heyminck}, {Hirota}, {Hoge}, {Huang}, {Impellizzeri},
  {Jiang}, {Kamble}, {Keisler}, {Kimura}, {Kono}, {Kubo}, {Kuroda}, {Lacasse},
  {Laing}, {Leitch}, {Li}, {Lin}, {Liu}, {Liu}, {Lu}, {Marson},
  {Martin-Cocher}, {Massingill}, {Matulonis}, {McColl}, {McWhirter}, {Messias},
  {Meyer-Zhao}, {Michalik}, {Monta{\~n}a}, {Montgomerie}, {Mora-Klein},
  {Muders}, {Nadolski}, {Navarro}, {Nguyen}, {Nishioka}, {Norton}, {Nystrom},
  {Ogawa}, {Oshiro}, {Oyama}, {Padin}, {Parsons}, {Paine}, {Pe{\~n}alver},
  {Phillips}, {Poirier}, {Pradel}, {Primiani}, {Raffin}, {Rahlin}, {Reiland},
  {Risacher}, {Ruiz}, {S{\'a}ez-Mada{\'\i}n}, {Sassella}, {Schellart}, {Shaw},
  {Silva}, {Shiokawa}, {Smith}, {Snow}, {Souccar}, {Sousa}, {Sridharan},
  {Srinivasan}, {Stahm}, {Stark}, {Story}, {Timmer}, {Vertatschitsch},
  {Walther}, {Wei}, {Whitehorn}, {Whitney}, {Woody}, {Wouterloot}, {Wright},
  {Yamaguchi}, {Yu}, {Zeballos}, and {Ziurys}]{m87_eht_II}
{Event Horizon Telescope Collaboration}.; {Akiyama}, K.; {Alberdi}, A.; {Alef},
  W.; {Asada}, K.; {Azulay}, R.; {Baczko}, A.K.; {Ball}, D.; {Balokovi{\'c}},
  M.; {Barrett}, J.;  et~al.
\newblock {First M87 Event Horizon Telescope Results. II. Array and
  Instrumentation}.
\newblock {\em \apjl} {\bf 2019}, {\em 875},~L2,
  \href{http://xxx.lanl.gov/abs/1906.11239}{{\normalfont
  [arXiv:astro-ph.IM/1906.11239]}}.
\newblock {\url{https://doi.org/10.3847/2041-8213/ab0c96}}.

\bibitem[{Shen} \em{et~al.}(2005){Shen}, {Lo}, {Liang}, {Ho}, and
  {Zhao}]{sgra_size}
{Shen}, Z.Q.; {Lo}, K.Y.; {Liang}, M.C.; {Ho}, P.T.P.; {Zhao}, J.H.
\newblock {A size of \~{}1AU for the radio source Sgr A* at the centre of the
  Milky Way}.
\newblock {\em \nat} {\bf 2005}, {\em 438},~62--64,
  \href{http://xxx.lanl.gov/abs/astro-ph/0512515}{{\normalfont
  [astro-ph/0512515]}}.
\newblock {\url{https://doi.org/10.1038/nature04205}}.

\bibitem[{Rioja} \em{et~al.}(2017){Rioja}, {Dodson}, {Orosz}, {Imai}, and
  {Frey}]{rioja_17}
{Rioja}, M.J.; {Dodson}, R.; {Orosz}, G.; {Imai}, H.; {Frey}, S.
\newblock {MultiView High Precision VLBI Astrometry at Low Frequencies}.
\newblock {\em \aj} {\bf 2017}, {\em 153},~105,
  \href{http://xxx.lanl.gov/abs/1612.02554}{{\normalfont
  [arXiv:astro-ph.IM/1612.02554]}}.
\newblock {\url{https://doi.org/10.3847/1538-3881/153/3/105}}.

\bibitem[{Hyland} \em{et~al.}(2022){Hyland}, {Reid}, {Ellingsen}, {Rioja},
  {Dodson}, {Orosz}, {Masson}, and {McCallum}]{hyland_22}
{Hyland}, L.J.; {Reid}, M.J.; {Ellingsen}, S.P.; {Rioja}, M.J.; {Dodson}, R.;
  {Orosz}, G.; {Masson}, C.R.; {McCallum}, J.M.
\newblock {Inverse Multiview. I. Multicalibrator Inverse Phase Referencing for
  Microarcsecond Very Long Baseline Interferometry Astrometry}.
\newblock {\em \apj} {\bf 2022}, {\em 932},~52,
  \href{http://xxx.lanl.gov/abs/2205.00092}{{\normalfont
  [arXiv:astro-ph.IM/2205.00092]}}.
\newblock {\url{https://doi.org/10.3847/1538-4357/ac6d5b}}.

\bibitem[{Reid} \em{et~al.}(2018){Reid}, {Loinard}, and {Maccarone}]{ngvla_lbs}
{Reid}, M.; {Loinard}, L.; {Maccarone}, T.
\newblock {Astrometry and Long Baseline Science}.
\newblock In Proceedings of the Science with a Next Generation Very Large
  Array; {Murphy}, E., Ed.,  2018, Vol. 517, {\em Astronomical Society of the
  Pacific Conference Series}, p. 523,
  \href{http://xxx.lanl.gov/abs/1810.06577}{{\normalfont
  [arXiv:astro-ph.GA/1810.06577]}}.

\bibitem[{Thompson} \em{et~al.}(2017){Thompson}, {Moran}, and {Swenson}]{TMSv3}
{Thompson}, A.R.; {Moran}, J.M.; {Swenson}, George~W., J.
\newblock {\em {Interferometry and Synthesis in Radio Astronomy}}, 3rd ed.;
  Springer: Cham,  2017.
\newblock {\url{https://doi.org/10.1007/978-3-319-44431-4}}.

\end{thebibliography}
\end{document}